\documentclass[a4paper,11pt]{article}

\usepackage{jheppub} 
\usepackage{subfigure}

\preprint{OU-HET-1064}

\title{\boldmath 
Exponential growth of out-of-time-order correlator without chaos: 
inverted harmonic oscillator
}

\author[a]{Koji Hashimoto,}
\affiliation[a]{Department of Physics, Osaka University,\\1-1 Machikaneyama, Toyonaka, Osaka 560-0043, Japan}
\author[b]{Kyoung-Bum Huh,}
\affiliation[b]{School of Physics and Chemistry, Gwangju Institute of Science and Technology, 123 Cheomdan- gwagiro, Gwangju 61005, Korea}
\author[b]{Keun-Young Kim,}
\author[a]{Ryota Watanabe}

\emailAdd{koji@phys.sci.osaka-u.ac.jp}
\emailAdd{hkabell1689@gist.ac.kr}
\emailAdd{fortoe@gist.ac.kr}
\emailAdd{watanabe@het.phys.sci.osaka-u.ac.jp}

\abstract{
We provide a detailed examination of a thermal out-of-time-order correlator (OTOC) growing exponentially in time in systems without chaos. The system is a one-dimensional
quantum mechanics with a potential whose part is an inverted harmonic oscillator.
We numerically observe the exponential growth of the OTOC when the temperature is higher than a certain threshold.
The Lyapunov exponent is found to be of the order of the classical Lyapunov exponent generated at the hilltop, and it remains non-vanishing even at high temperature.
We adopt various shape of the potential and find these features universal.
The study confirms that the exponential growth of the thermal OTOC does not necessarily mean chaos when the potential includes a local maximum.
We also provide a bound for the Lyapunov exponent of the thermal OTOC
in generic quantum mechanics in one dimension, which is of the same form as the chaos bound 
obtained by Maldacena, Shenker and Stanford.
}

\begin{document}

\maketitle
\flushbottom

\section{Introduction}
\label{sec:intro}

The exponential growth of out-of-time-order correlator (OTOC) \cite{Larkin} has attracted considerable attention these years, motivated by possible relations between black hole systems and quantum mechanical systems through the AdS/CFT correspondence \cite{Maldacena:1997re}. 
The ``chaos bound'' \cite{Maldacena:2015waa} for the Lyapunov exponent $\lambda_\text{OTOC}$ in thermal OTOCs in large $N$ quantum theories 
at temperature $T$,
\begin{align}
    \lambda_\text{OTOC}(T) \leq 2\pi T \, ,
    \label{MSSbound}
\end{align}
is saturated when there exists a gravity dual in which the
Lyapunov exponent is interpreted as a red shift factor near the black hole horizon probed by shock waves \cite{Shenker:2013pqa,Shenker:2013yza,Leichenauer:2014nxa}. This indicator of the holographic principle indeed has lead \cite{Kitaev-talk-KITP,Maldacena:2016hyu} to a surprising quantum mechanical model, the Sachdev-Ye-Kitaev (SYK) model \cite{Sachdev:1992fk,Kitaev-talk}, which admits a 2-dimensional dual gravity description.

With the OTOC as the novel indicator of quantum chaos, quantum chaotic few-body systems have been probed to see whether the OTOC grows exponentially in time. 
The way to calculate microcanonical/thermal OTOCs in generic quantum mechanics was provided \cite{Hashimoto:2017oit}, and 
major examples of chaotic systems with the exponentially growing OTOCs include
a kicked rotor \cite{Rozenbaum:2017zfo}, a stadium billiard \cite{Hashimoto:2017oit,Rozenbaum:2019kdl}, the Dicke model \cite{Chavez-Carlos:2018ijc}, bipartite systems \cite{Prakash:2020fsj,Prakash:2019kip}, and coupled harmonic oscillators \cite{Akutagawa:2020qbj}\footnote{See \cite{Zhuang:2019jyq} for the OTOC analysis for the Henon-Heiles system. Various kinds of OTOCs in quantum maps were also studied \cite{Garcia-Mata:2018slr,Lakshminarayan:2019xxd,Bergamasco:2019tfw,Fortes:2019frf}.
The cases with large $N$
are found in \cite{Shen:2016htm,Bohrdt:2016vhv,Bianchi:2017kgb,Lin:2018tce,Rammensee:2018pyk,Lin:2018luj,Wang:2019vjl,Hartmann:2019cxq,Dag:2019yqu,Borgonovi:2019mrk,Ghosh:2019yjh,Yan:2019wio}.}. 
In particular, in the coupled harmonic oscillator system \cite{Akutagawa:2020qbj} 
(which is reminiscent of Yang-Mills theory
\cite{Matinyan Savvidy Ter-Arutunian Savvidy (1981a), Matinyan Savvidy Ter-Arutunian Savvidy (1981b), Savvidy (1984)}), 
the thermal OTOC is a better indicator of quantum chaos compared to the conventional energy level statistics.

The important observation was made in \cite{Pappalardi:2018frz,Hummel:2018brt,Pilatowsky-Cameo:2019qxt} finding that the exponential growth of OTOCs is possible in non-chaotic regular systems at low dimensions\footnote{See also related discussions in \cite{Rozenbaum:2019nwn,Li:2020zuj}.}. This growth is interpreted as being generated by a classical unstable maximum of the potential at which, locally, an initial difference grows exponentially in time. The phenomenon is expected to be general, and
\cite{Xu:2019lhc} provided a general semiclassical inequality between the classical Lyapunov exponent $\lambda_\text{saddle}$ at the unstable maximum (or a saddle point) and the quantum Lyapunov exponent $\lambda_\text{OTOC}$ of the thermal OTOC at infinite temperature,
\begin{align}
    \lambda_\text{OTOC}(T=\infty)\geq \lambda_\text{saddle} \, .
\label{boundsaddle}
\end{align}
Since a classical saddle or a local maximum of the potential does not necessarily mean classical chaos, this inequality \eqref{boundsaddle} suggests that the information scrambling is not only generated by chaos, and that the scrambling is possible in regular systems.

The two inequalities, \eqref{MSSbound} and \eqref{boundsaddle}, 
inevitably lead us to the following two questions: $\langle i \rangle$ {\it whether the general inequality \eqref{boundsaddle} applies to any quantum mechanics or not}, and $\langle ii \rangle$ {\it what is the relation between \eqref{MSSbound} and \eqref{boundsaddle}}. We are going to study these two questions in this paper. 

Concerning the first question $\langle i \rangle$, we examine OTOCs for the system of one-dimensional inverted harmonic oscillator. In one dimension, this is the most generic set-up which generates a non-zero positive classical Lyapunov exponent $\lambda_\text{saddle}$ at the local maximum. To make the system bounded from below to define the temperature $T$, we put some potential walls away from the local maximum. This well-defined quantum system of a double-well potential is non-chaotic due to the Poincar\'e-Bendixon theorem. We numerically calculate the thermal OTOCs for various types of the walls and at various values of $T$. We indeed find a nonzero $\lambda_\text{OTOC}$, so, we confirm that in generic one-dimensional quantum mechanics with a local maximum in the potential, in spite of the non-chaoticity, the OTOCs grow exponentially in time. The observed $\lambda_\text{OTOC}$ is $T$-dependent, and its value is ${\cal O}(\lambda_\text{saddle})$, 
thus naturally interpreted as being generated by the inverted harmonic potential. The infinite temperature limit of $\lambda_\text{OTOC}(T)$ slightly violates \eqref{boundsaddle}, which would be due to our fully quantum calculations away from the semiclassical limit.

As an answer to the second question $\langle ii \rangle$, we derive an inequality 
\begin{align}
    \lambda_\text{OTOC}(T) \leq c\,  T \, , \quad c \simeq {\cal O}(1)
    \label{bound}
\end{align}
for generic one-dimensional quantum mechanics. In fact, the structure of the inverted harmonic oscillator potential, together with the quantum resolution condition to discriminate the local maximum by wave functions, leads to this inequality. The similarity to the ``chaos bound'' \eqref{MSSbound} is striking. The bound \eqref{MSSbound} is for large $N$ theories while our inequality \eqref{bound} is for a single degree of freedom. 

A possible discussion to relate \eqref{MSSbound} with \eqref{bound} owes to AdS/CFT set-ups. The renowned large-$N$ quantum mechanics with a dual gravity description is the BFSS matrix theory \cite{Banks:1996vh}\footnote{This was a motivation for the model of \cite{Akutagawa:2020qbj},
and string theory matrix models in similar spirit are found in  \cite{Asano:2015eha, Hashimoto:2016wme, Berenstein:2016zgj, Akutagawa:2018yoe}.}. Separating one degree of freedom and integrating out the remaining as a black hole \cite{Iizuka:2001cw}, the system reduces to a quantum mechanics of a particle in one dimension\footnote{For related chaos analyses, see 
\cite{Gur-Ari:2015rcq,Berkowitz:2016znt,Buividovich:2018scl}.}. 
This particle feels the gravitational potential emergent from the integration. 
It is known that there is a universal chaotic behavior near black hole horizons \cite{Hashimoto:2016dfz}\footnote{The potential
provides a way to explain Hawking radiation and other universal phenomena 
\cite{Betzios:2016yaq, Morita:2018sen, Dalui:2018qqv, Hashimoto:2018fkb, Zhao:2018wkl, Morita:2019bfr}.} which is due to the inverted harmonic (gravitational) potential with the Lyapunov exponent $2\pi T$. Therefore, proper integration of large degrees of freedom, in a quantum mechanics with a gravity dual, may lead to an effective one-dimensional quantum mechanics with the inverted harmonic potential. Although this whole story is still far from our reach, it motivates us to the study given in this paper and to provide the answers to the two questions $\langle i \rangle$ $\langle ii \rangle$ described above.

This paper is organized as follows. 
In Sec.~\ref{sec:2}, we calculate the thermal OTOC in the quantum mechanics of the simplest inverted harmonic oscillator (a double-well Higgs-like potential). We find the temperature-dependent quantum Lyapunov exponents, whose high temperature limit remains nonzero.
In Sec.~\ref{sec:3}, we study the universality of the exponential growth of
the thermal OTOC, by evaluating quantum models with a different shape of the potential walls.
In Sec.~\ref{sec:4}, we derive the inequality that the Lyapunov exponent of the thermal OTOC is bounded above by the temperature, in a generic one-dimensional quantum mechanics. 
Sec.~\ref{sec:5} is for our summary and discussions.

Note added: while we were finishing our project, we noticed a related paper
\cite{Bhattacharyya:2020art} which studies an OTOC for a system with an 
inverted harmonic oscillator.


\section{Exponential growth of OTOC in inverted harmonic oscillator}
\label{sec:2}
In this section we study the microcanonical and thermal OTOCs of the simplest quantum mechanical system including an inverted harmonic oscillator (IHO). 
We employ a one-dimensional Hamiltonian system, which is hence classically non-chaotic (regular), while at the unstable maximum of the potential, a nonzero Lyapunov exponent $\lambda_\text{saddle}$ appears. 
We numerically find the microcanonical/thermal OTOCs grow exponentially at early times. We study the temperature dependence of the observed quantum Lyapunov exponents $\lambda_\text{OTOC}$ of the thermal OTOCs and find that at the high temperature limit the Lyapunov exponent $\lambda_\text{OTOC}$ remains nonvanishing, whose value is ${\cal O}(\lambda_\text{saddle})$.

The simplest quantum mechanical system including the inverted harmonic oscillator is defined by the Hamiltonian
\begin{eqnarray}
\label{eq:IHO}
	& H \equiv p^2 + V\,, \\
	& V \equiv g \left(x^2 - \displaystyle\frac{\lambda^2}{8g}\right)^2      
	= - \frac{1}{4}\lambda^2 x^2 + g x^4 + \displaystyle\frac{\lambda^4}{64g}\,.
	\label{HP}
\end{eqnarray}
Here $\lambda$ and $g$ are constant parameters. This is nothing but the Higgs potential in the high energy theoretic terminology.
The $x^4$ term is included in order for the system to be bounded from below\footnote{The boundedness of the potential ensured by the ``soft wall'' $x^4$ term is necessary to define temperature which is crucial to our analyses. It was pointed out in \cite{Ali:2019zcj} that using an analytic continuation of the frequency parameter $\omega$, the thermal OTOC for the standard harmonic oscillator $c_n(t) = \cos^2 \omega t$ obtained in \cite{Hashimoto:2017oit} suggests $c_n(t) = \cosh^2 \omega t$ for
an inverted harmonic oscillator without the boundedness. This grows exponentially forever. However, a naive analytic continuation makes the energy to be pure imaginary, and the definition of the microcanonical/thermal OTOCs are ambiguous. To define them properly with the temperature, we consider only the cases with bounded potentials by introducing the walls. See Sec.~\ref{sec:3} for the dependence of the choice of the walls. And in Sec.~\ref{sec:4}, the bounded bottom of the potential is indeed crucial for the derivation of the inequality \eqref{boundT}. }. 

Since this is a one-dimensional Hamiltonian system, the classical mechanics is regular. But this does not mean that the Lyapunov exponent vanishes. The system is unstable around $x=0$, so we have a non-vanishing classical Lyapunov exponent $\lambda_\text{saddle}$ there. This exponent is equal to the parameter $\lambda$ in the potential given above. Note that the parameter $\lambda$ determines the curvature of the unstable top of the hill. 
In this section first we choose $\lambda=2$ and $g=1/50$ for our numerical calculations of the OTOCs, and later we choose another set $\lambda=2\sqrt{5}$ and $g=1/10$. The latter shares, with the former, the property that the location of the bottom of the potential is at $x=\pm 5$\footnote{By rescaling $x$ and $H$, we can tune a certain combination of $\lambda$ and $g$ to be an arbitrary value, for example, $\lambda/\sqrt{8g}=5$.}.

\begin{figure}
\centering
     \subfigure[Potential shape and energy levels.]
     {\includegraphics[width=6cm]{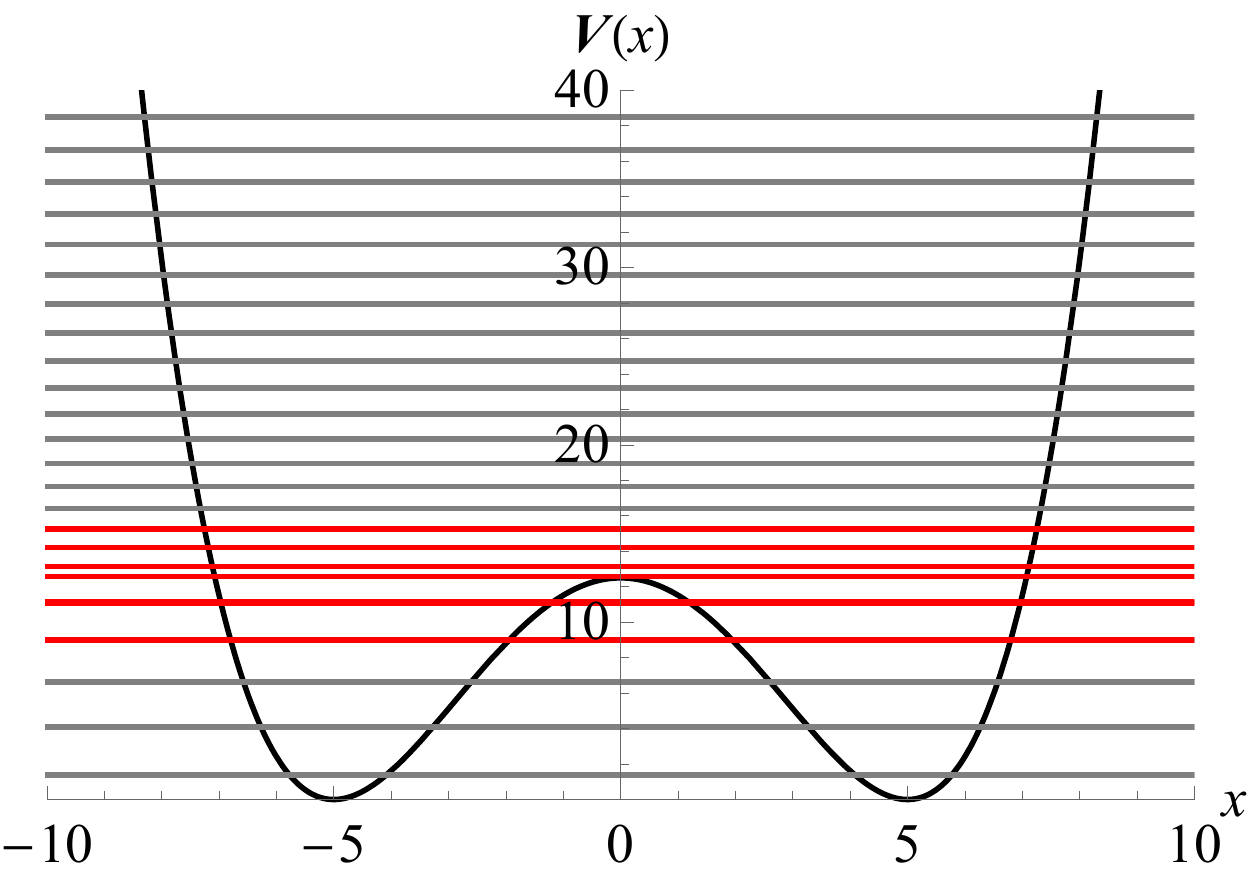} \label{fig1a}}
     \subfigure[Energy eigenvalues]
     {\includegraphics[width=6cm]{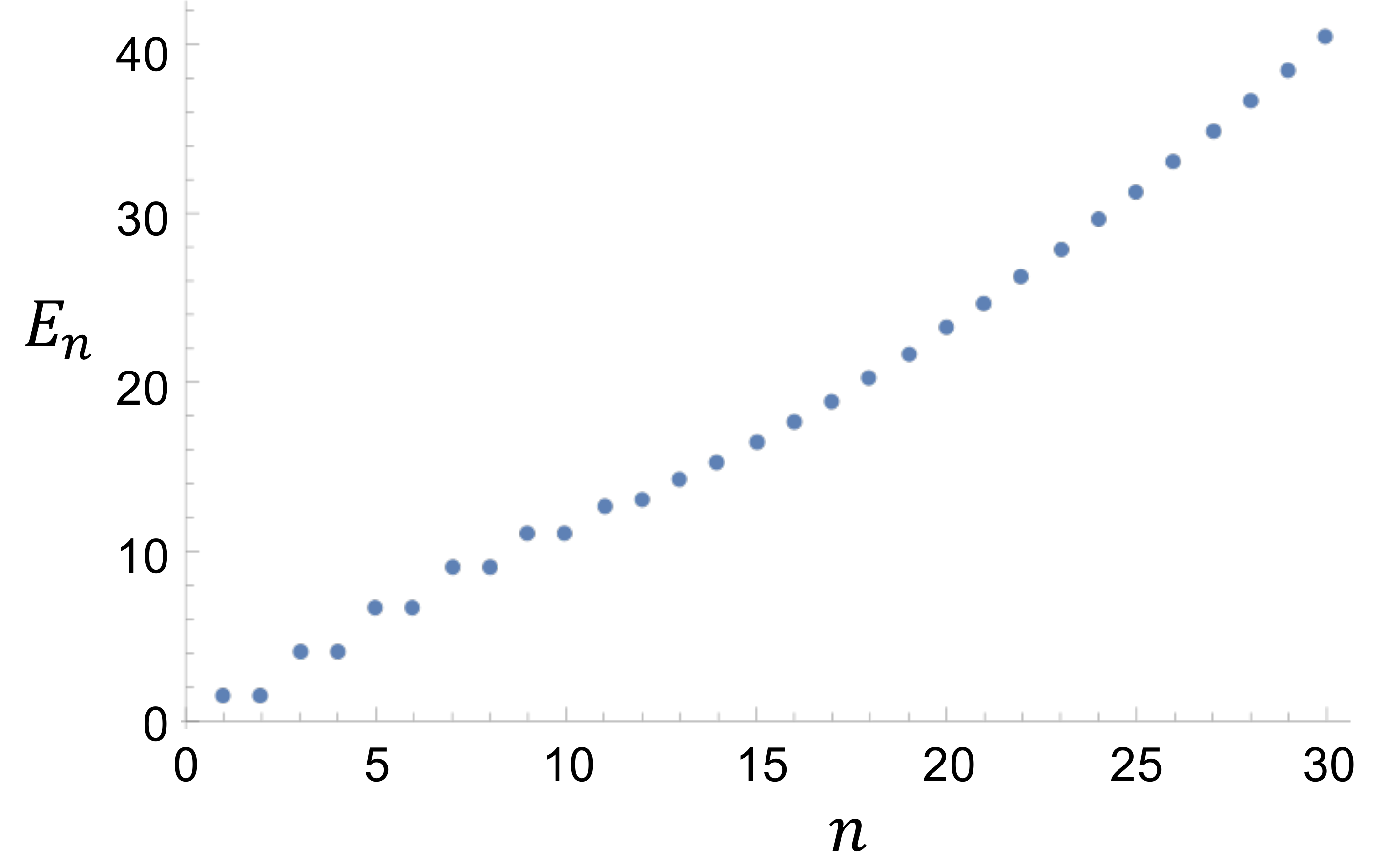} }
 \caption{Inverted harmonic oscillator potential for $\lambda=2, g=1/50$ ($V(0)=12.5$)   and its energy levels. The energy levels in red color play an important role for exponential growth of OTOCs. See Fig.~\ref{fig:microcanonical}. Note that the energy levels smaller than $n=11$ are almost degenerated so the black lines below the top of the hill in (a) are double lines. }\label{fig:energy_eigenvalues}
\end{figure}

We are interested in the quantum analogue of the exponential behavior of the particle motion around the top of the hill, thus we choose the following 
thermal OTOC defined in the Heisenberg picture\footnote{
In this paper we use only the OTOC with the commutator squared, to make the story parallel to the classical definition of chaos and the Lyapunov exponent. For a more general OTOC without the commutator squared is 
studied in Appendix \ref{App:B}.
} 
by 
\begin{equation}
\label{eq:thermal OTOC}
	C_T(t) \equiv -\langle [x(t),p(0)]^2 \rangle,
\end{equation}
where $\langle O \rangle \equiv {\rm tr}[e^{-\beta H}O]/{\rm tr}[e^{-\beta H}]$ is the thermal average. Let $|n\rangle$ be the $n$-th energy eigenstate, $H|n\rangle=E_n|n\rangle$ ($n=1,2,3,\cdots$). We define the microcanonical OTOC for this energy eigenstate by
\begin{equation}
\label{eq:microcanonical OTOC}
	c_n(t) \equiv -\langle n | [x(t),p(0)]^2 | n \rangle\, .
\end{equation}
Using the completeness relation of the energy eigenstates, the thermal OTOC can be written as the thermal average of the microcanonical OTOCs,
\begin{equation}
\label{eq:thermal averave}
	C_T(t) = \frac{1}{Z}\sum_n e^{-\beta E_n}c_n(t)\, , \quad Z \equiv {\rm tr}[e^{-\beta H}]\, .
\end{equation}

We quantize the system and consider the time-independent Schr\"odinger equation
\begin{equation}
\label{eq:Schrodinger IHO}
	-\frac{d^2}{dx^2}\psi_n(x) + \left[ - \frac{1}{4}\lambda^2 x^2 + g x^4 + \frac{\lambda^4}{64g} \right]\psi_n(x) = E_n\psi_n(x)\, , 
\end{equation}
where we take $\lambda=2, g=1/50$. We numerically solve this equation and obtaine energy eigenvalues $E_n$ and the wave functions $\psi_n(x)$. In Fig.\ref{fig:energy_eigenvalues}, we show the obtained distribution of the energy eigenvalues.

\begin{figure}
	\centering
    \includegraphics[width=90mm]{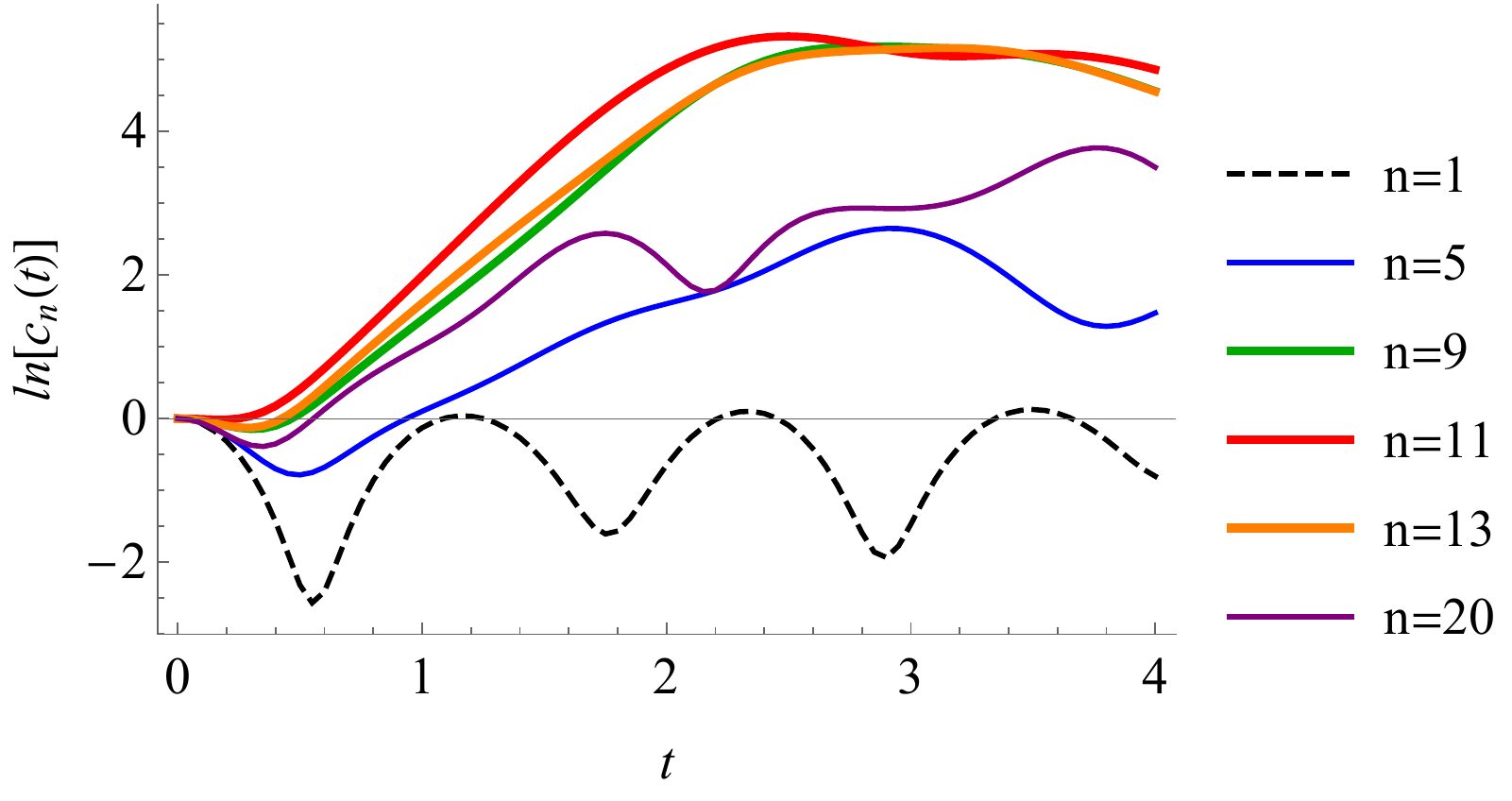}
	\caption{The microcanonical OTOCs for the IHO. We can see the strong exponential growth for intermediate modes (like $n=9\sim 13$), while lower modes and higher modes do not show initial exponential growth. The energy range of these intermediate modes correspond to the local maximum of the potential (the red lines in Fig.\ref{fig1a}).}
	\label{fig:microcanonical}
\end{figure}

Following the general method for calculating the OTOCs numerically \cite{Hashimoto:2017oit}, we compute\footnote{In the evaluation, we include the energy eigenstates up to $n=192$.} the microcanonical OTOCs as functions of $t$ for each energy level $n$. In Fig.\ref{fig:microcanonical}, we show our numerical results. For lower/higher modes, the OTOCs do not show exponential growth. On the other hand, for intermediate modes ($n=9,10,11,12,13$), the OTOCs exponentially grow at early times. These intermediate energy eigenvalues are in the range $8<E<14$. Actually, the height of the unstable saddle from the bottom of the potential is $\frac{\lambda^4}{64g}=12.5$. 
Note that the level $n=11$ (red in Fig.\ref{fig:microcanonical} ) is the closest\footnote{The energy levels below the top of the hill in Fig.\ref{fig1a} are double lines.} to the top of the potential (Fig.\ref{fig1a}) and shows the strongest exponential growth.

\begin{figure}
	\centering
    \includegraphics[width=90mm]{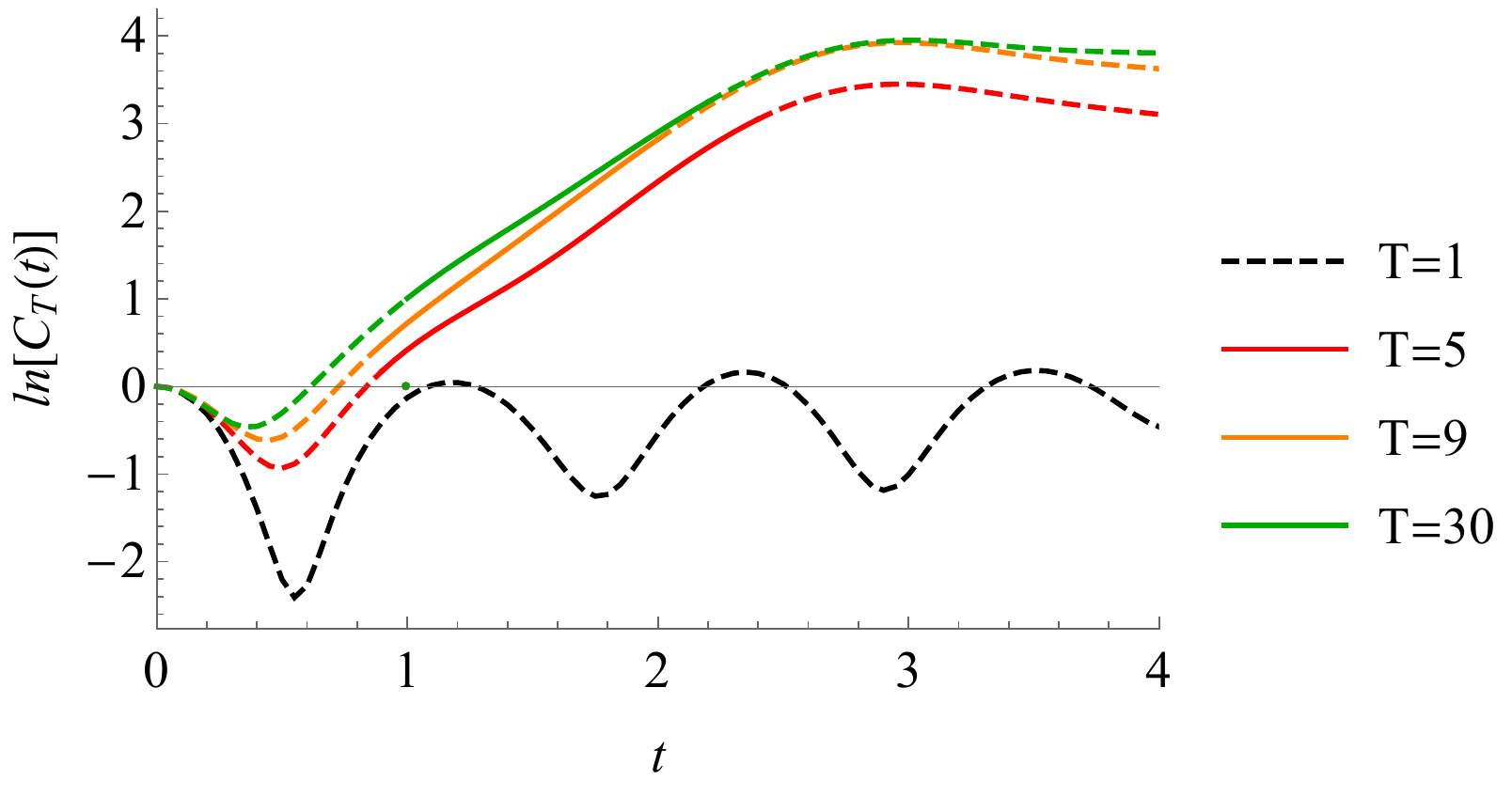}
	\caption{The time dependence of the thermal OTOCs of the system \eqref{HP} for various values of the temperature $T$. The dashed part is non-linear and the solid part is linear (exponential in $t$). The time domains for the solid lines are determined at each temperature data such that the linear fit provides the smallest confidence interval of the slope normalized by the slope itself. The obtained time domains are longer than the twice of ${\cal O}(1/\lambda_\text{OTOC})$, which certifies the exponential growth here.
	}
	\label{fig:thermal_IHO}
\end{figure}

The behavior of the microcanonical OTOCs is exactly what we expect from the structure of the IHO potential. When the energy is low, the wave function lives inside the well and do not reach the unstable saddle. If we raise the energy, the wave function begin to feel the hilltop of the potential. In addition, the wave function localizes around the unstable point\footnote{By the conservation of energy, the momentum of a particle is small when the potential is high. This means that the particle stays for a longer time around the turning point and the hilltop of the potential. Quantum mechanically, this means the wave function localizes around those points. We would like to thank Lea Ferreira dos Santos and Sa\'ul Pilatowsky for valuable discussions on localization of wave functions.}. 
As a result, the corresponding microcanonical OTOC shows the exponential growth. When the energy is high enough, the effect of the unstable point on the wave function is buried, and the corresponding OTOCs do not show the exponential growth any more. In this IHO case, the origin of the exponential growth of the microcanonical OTOC is not chaoticity, but instability of the potential. Hence, the exponential growth of the OTOC does not necessarily indicate chaos.

By taking the thermal average of the microcanonical OTOCs, we compute the thermal OTOCs for various values of the temperature\footnote{The computation was done by discretizing the time coordinate by units of $0.1$. In Fig.~\ref{fig:thermal_IHO}, we connected those discrete points by smooth curves for a better visibility.}. The numerical results are shown in Fig.~\ref{fig:thermal_IHO}. We can find the exponential growth in the thermal OTOCs for high temperature\footnote{Interestingly, the Ehrenfest time looks almost the same (around $t=3$) for this region of $T$.}. 

\begin{figure}
	\centering
	\includegraphics[width=90mm]{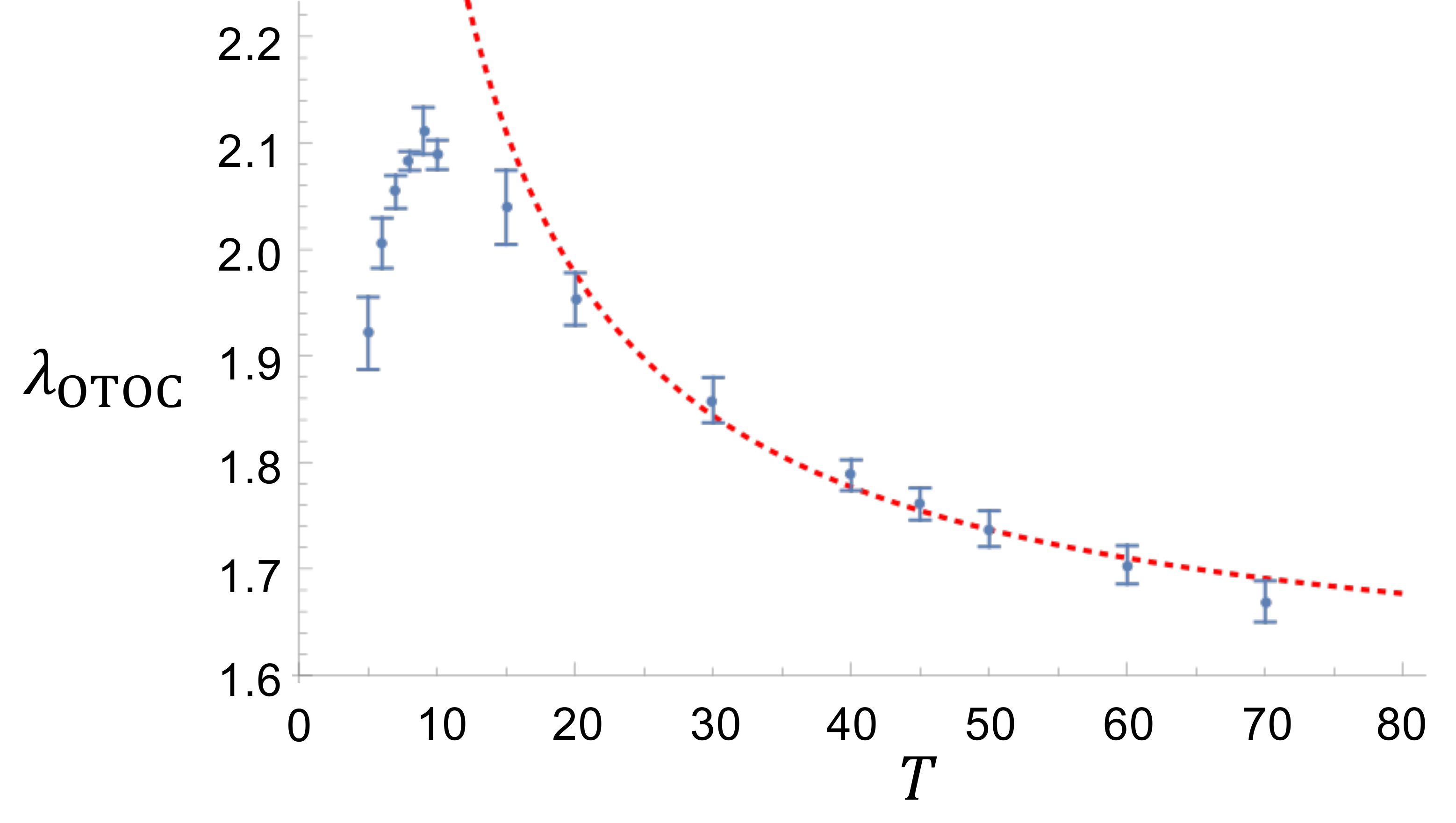}
	\caption{The temperature dependence of the Lyapunov exponents $\lambda_\text{OTOC}$ of the thermal OTOCs. The bar represents the 95\% confidence interval for the exponent. The dashed curve is the fitting function for the Lyapunov exponents obtained in the range $20\leq T \leq 70$.}
	\label{fig:Lyapunov_IHO}
\end{figure}

We fit the thermal OTOCs at early times by a function $a \exp[\lambda_{\rm OTOC}\, t]$ with $a$ and $\lambda_\text{OTOC}$ being adjustable constant parameters, to find the Lyapunov exponents $\lambda_{\rm OTOC}$. 
In other words, the slope of the solid part in Fig.~\ref{fig:thermal_IHO} is the Lyapunov exponent at a given value of the temperature.
The temperature dependence of the Lyapunov exponents is shown in Fig.~\ref{fig:Lyapunov_IHO}. The bars represent the 95\% confidence interval for the Lyapunov exponent at a given value of the temperature.

Here, from the obtained Lyapunov exponents, we observe the following facts. First, the order of those measure exponents is equal to that of the classical Lyapunov exponent $\lambda_\text{saddle}=2$.
Thus, we find that indeed the classical instability of the unstable maximum of the IHO is detected by the thermal OTOC\footnote{The OTOC is a quantum counterpart of the square of the classical Poisson bracket. In view of this, in the comparison between $\lambda_\text{OTOC}$ and $\lambda_\text{saddle}$, a natural relation would be $\lambda_{\rm OTOC} = 2 \lambda_\text{saddle}$. However, considering a compression factor of the phase space for the dominantly growing mode \cite{Xu:2019lhc}, this factor $2$ may drop off. 
See \cite{Xu:2019lhc} for the discussion.}.

Second, 
as the temperature goes up, the exponent slightly decreases monotonically. To check if the inequality \eqref{boundsaddle} is satisfied in our IHO system, we study the Lyapunov exponent in the high-temperature limit $T \to \infty$. To see this, we assume that $\lambda_{\rm OTOC}(T)$ is analytic around $T=\infty$, that is, it can be expanded as
\begin{equation}
\label{eq:expansion}
\lambda_{\rm OTOC}(T)=a_0+\frac{a_1}{T}+\frac{a_2}{T^2}+\frac{a_3}{T^3}+\cdots \, .
\end{equation}
Using this as a fitting function, we find
\begin{equation}
\label{eq:fitting}
	\lambda_{\rm OTOC}(T) \sim 1.58+\frac{8.01}{T}\, .
\end{equation}
as a reasonable fitting function for $\lambda_\text{OTOC}$ in the high temperature region. 
In Fig.~\ref{fig:Lyapunov_IHO}, this fitting function is drawn as the dashed curve. In Appendix \ref{App:A}, we discuss the error analysis for the fitting. The fitting function \eqref{eq:fitting} is non-negative, which satisfies the general requirement that the Lyapunov exponent is non-negative by definition.

\begin{figure}
\centering
     \subfigure[Potential shape and energy levels.  The energy levels in red color play an important role for exponential growth of OTOCs.]
     {\includegraphics[width=6cm]{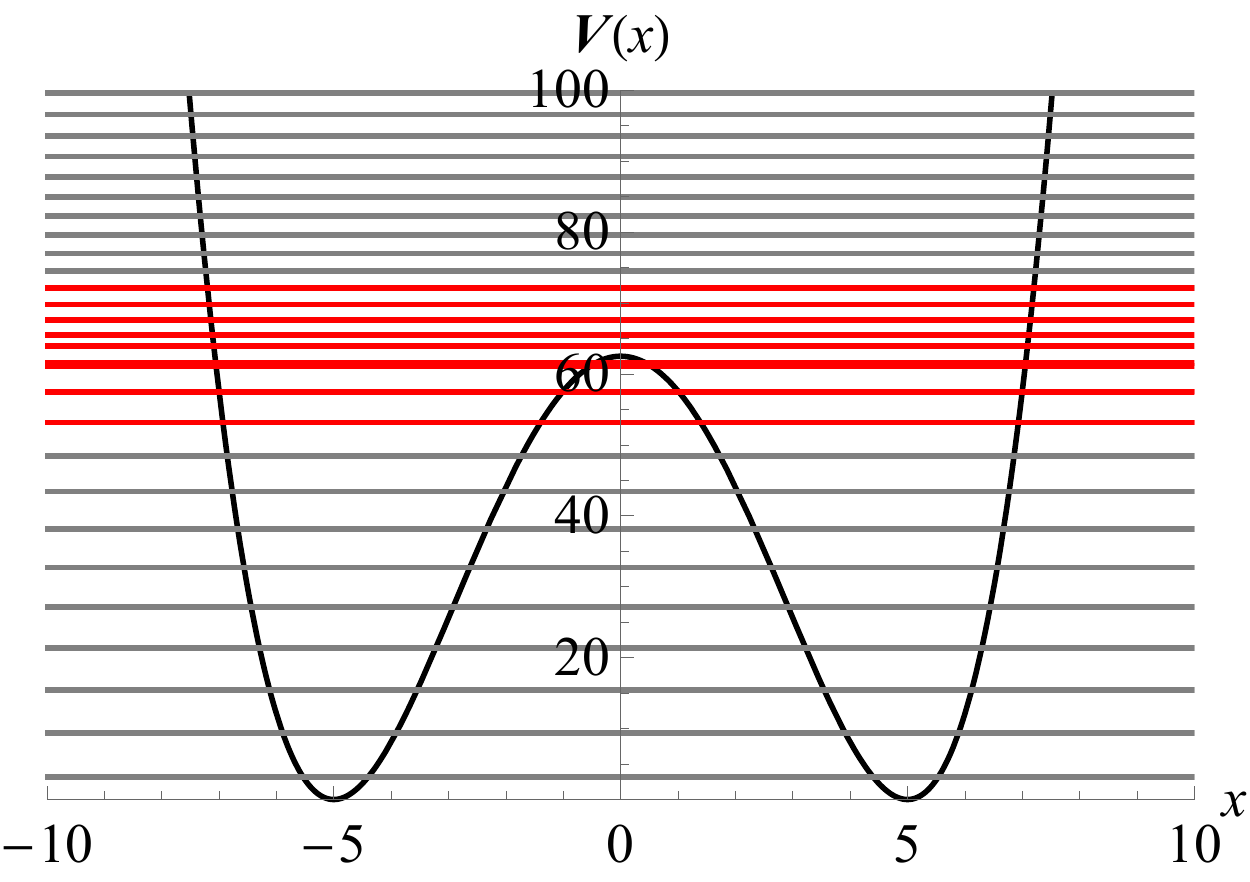} \label{pot2a}} \ \ \
     \subfigure[Quantum Lyapunov exponent vs temperature. The red fitting curve is $\lambda_{\rm OTOC}(T) \sim 4.01+35.4/T$.]
     {\includegraphics[width=6cm]{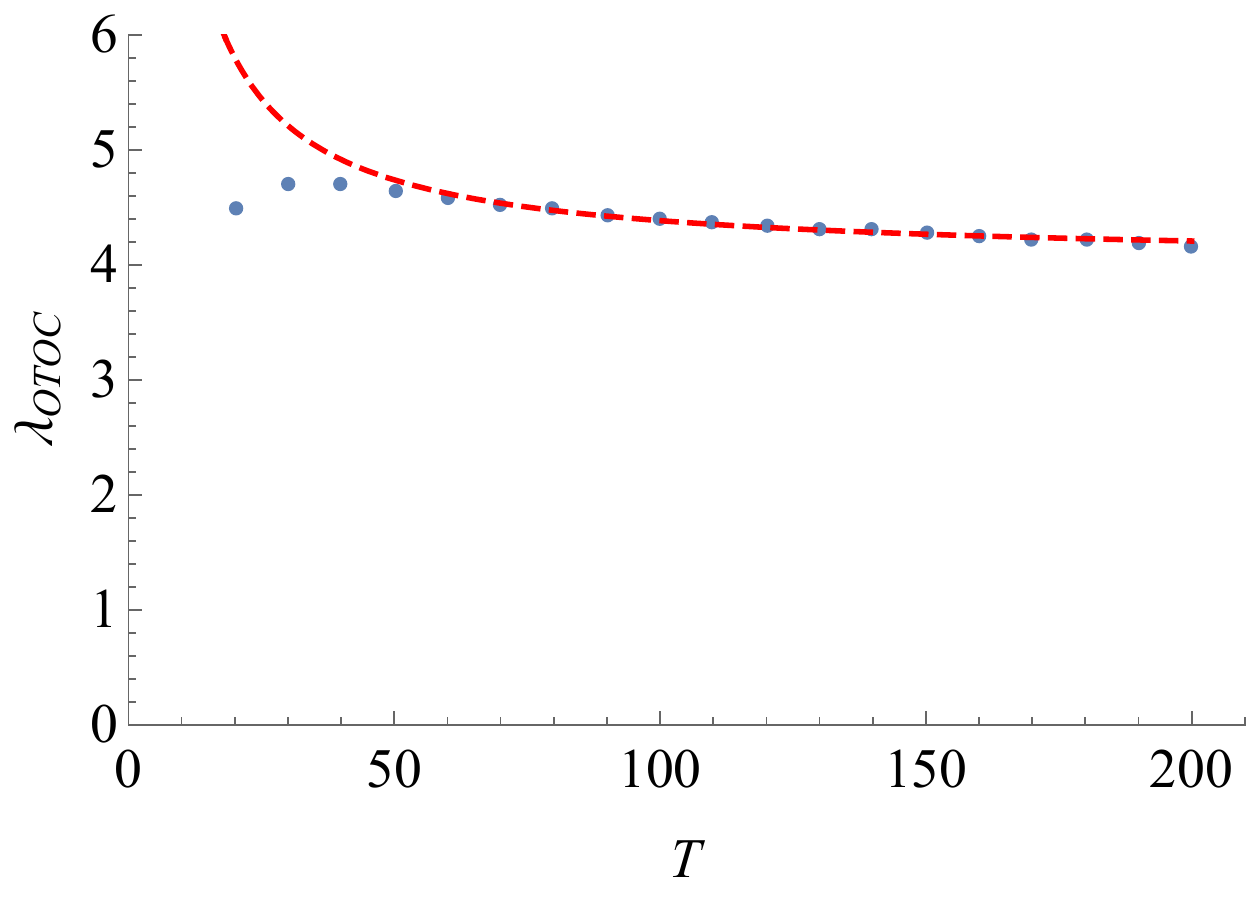} \label{pot2b}}
 \caption{Inverted harmonic oscillator potential for $\lambda =2\sqrt{5}, g = 1/10, (V(0) = 62.5)$.}\label{pot2}
\end{figure}

We repeat the analysis for another potential with parameters $\lambda = 2\sqrt{5}$ and $g = 1/10$, which are chosen such that the potential minimum is located at $x=\pm 5$ for the comparison with the previous case.  See Fig.~\ref{pot2a} for the potential shape, where the energy levels are displayed as horizontal lines. Similarly to the previous case in Fig.~\ref{fig:energy_eigenvalues}, the states with energy levels around the local maximum play an important role. Only the microcanonical canonical OTOCs of those ``red'' states show the exponential growth at early times. Furthermore, these are dominant contributions to the exponential growth of the thermal OTOCs. 
Fig.~\ref{pot2b} shows the thermal Lyapunov exponents as a function of temperature. The high temperature region is fitted by the red dashed line:
\begin{equation}
\label{eq:fitting2}
	\lambda_{\rm OTOC}(T) \sim 4.01+\frac{35.4}{T}\, .
\end{equation}

Importantly, the Lyapunov exponent does not vanish in the high-temperature limit. 
The asymptotic value of \eqref{eq:fitting} and \eqref{eq:fitting2} at $T=\infty$ is smaller than the classical Lyapunov exponent ($\lambda_{\mathrm{saddle}}=2$ for the first case and $\lambda_{\mathrm{saddle}}=2\sqrt{5} \sim 4.47$ for the second case). This appears to slightly violate the proposed inequality \eqref{boundsaddle}.
Noting that the inequality \eqref{boundsaddle} was derived in the classical limit \cite{Xu:2019lhc}, this slight violation would be due to the quantum effect. In addition, the Hilbert space of our quantum mechanical system is infinite dimensional, thus the infinite temperature limit is not well-defined.
These would be possible reasons for the slight violation of the inequality \eqref{boundsaddle}. Nevertheless, the observation that the quantum Lyapunov exponent $\lambda_\text{OTOC}$ asymptotes to a nonzero constant at $T=\infty$ is one of our important conclusions.

The measured $\lambda_\text{OTOC}$ is a monotonicaly decreasing function of $T$. This can be naturally understood as follows. 
If we raise the temperature, the higher modes of the microcanonical OTOCs contribute to the thermal OTOC. In the IHO system, since the microcanonical OTOCs for the higher modes do not show any exponential growth, they do not contribute to the exponential behavior of the thermal OTOC, rather may smear it. Hence, $\lambda_{\rm OTOC}(T)$ is expected to be a monotonically decreasing function of $T$, $d\lambda_{\rm OTOC}/dT\leq0$. This is explicitly observed in our numerical evaluation of $\lambda_\text{OTOC}(T)$.
However, this is not a universal feature because there are also cases where $d\lambda_{\rm OTOC}/dT\geq0$. It depends on the shape of potential as will be shown in the following section.

In this section, we have taken the simplest potentials which include the inverted harmonic oscillator, and have seen that the Lyapunov exponent of the thermal OTOCs is non-vanishing. This explicitly shows that the thermal OTOCs can grow exponentially even in non-chaotic systems. A possible concern would be that this result may be specific to the Higgs-type potential \eqref{HP}. To resolve the issue, in the next section we shall see the universality of the results.


\section{Universality of the growth}
\label{sec:3}

\begin{figure}
\centering
     \subfigure[$\lambda =2\sqrt{5}, g = 1/10, V(0) = 62.5$]
     {\includegraphics[width=6cm]{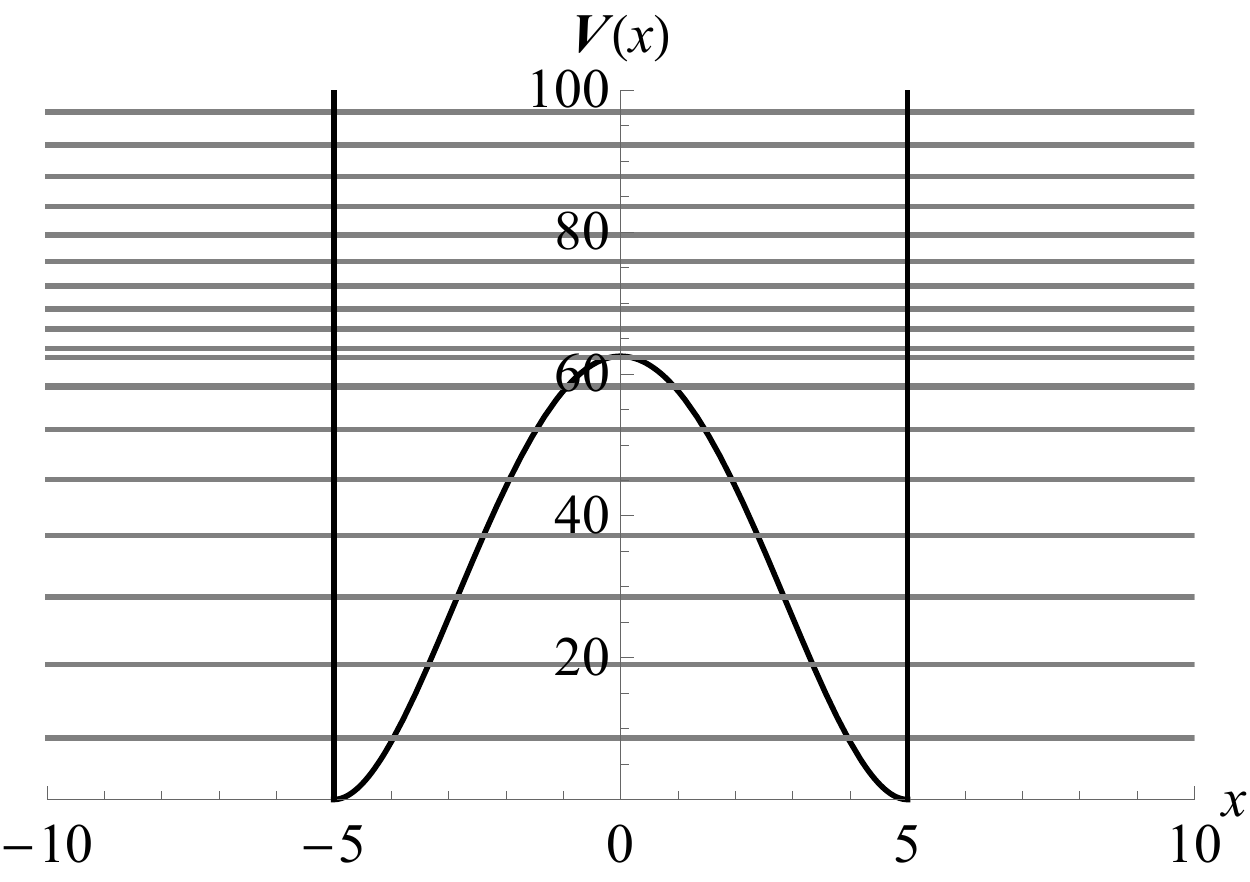} \label{pothhha}} \ \ \
     \subfigure[$\lambda =2, g = 1/50, V(0)= 12.5$]
     {\includegraphics[width=6cm]{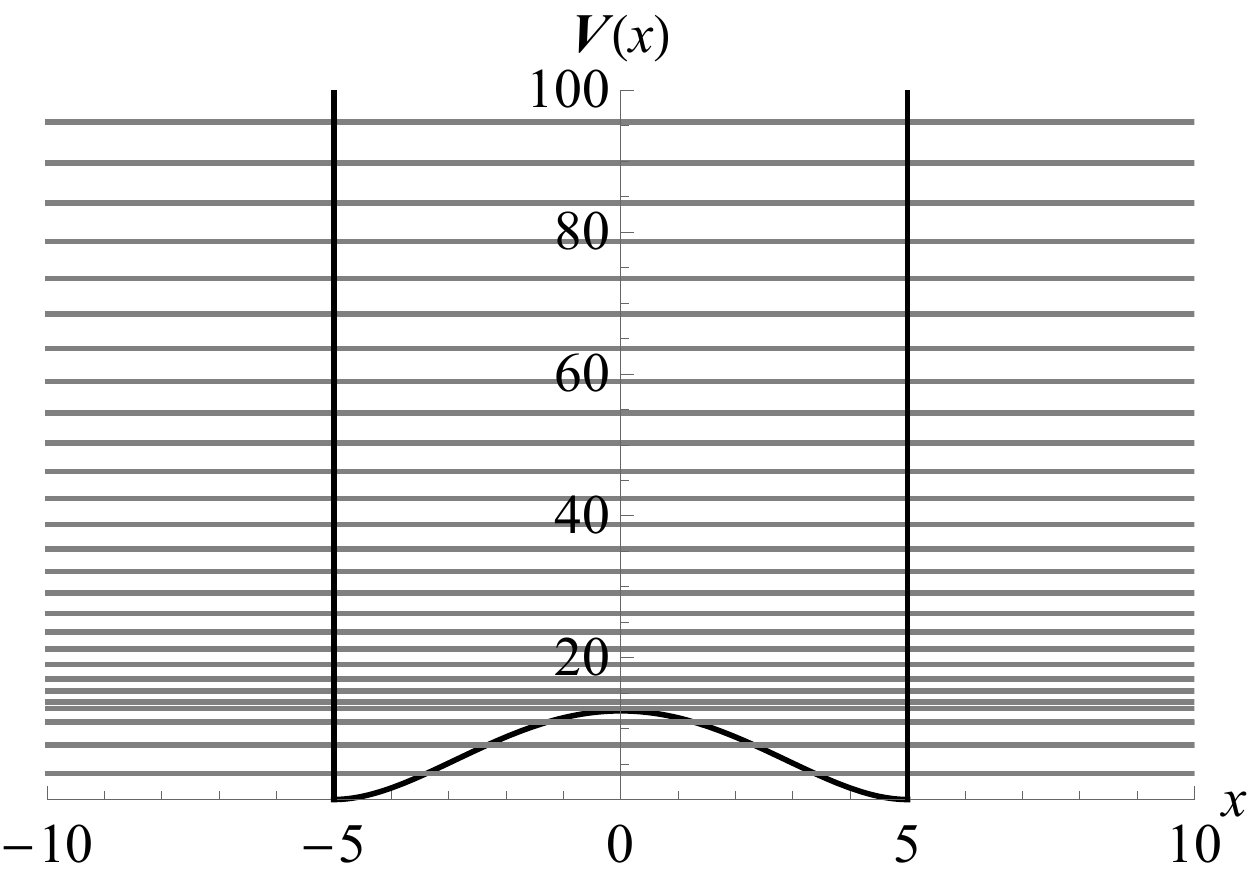} \label{pothhhb}}
 \caption{Potential energy with hard walls and energy levels. The levels below the top of the hill are double lines.}\label{pothhh}
\end{figure}

In this section, 
we study the universality of the exponential growth phenomenon of the thermal OTOC in one-dimensional quantum mechanics. In the previous section, we have used the
potential of the Higgs-type \eqref{HP}. To study the universality, let us consider the following potential:
\begin{align}
V(x) = \left\{
\begin{array}{lll}
\left(x^2 - \displaystyle\frac{\lambda^2}{8g}\right)^2      
	= - \frac{1}{4}\lambda^2 x^2 + g x^4 + \displaystyle\frac{\lambda^4}{64g} &\quad & \left(|x| \leq \frac{\lambda}{\sqrt{8g}}\right)\\
\infty & \quad & \left(\frac{\lambda}{\sqrt{8g}} < |x|\right)
\end{array}
\right.
\label{Vhardd00}
\end{align} 
This potential shares the same form as \eqref{HP} inside, but we put hard walls at $x = \pm\frac{\lambda}{\sqrt{8g}} $.
See, for example, Fig.~\ref{pothhh} for the shape of the potentials with the same values of the parameters as the ones in Sec.~\ref{sec:2}.

We investigate this hard-wall model for two reasons. First, 
since it shares the same potential inside as that of Sec.~\ref{sec:2}, thus, while the classical saddle effect is kept, the effect of the boundaries can be efficiently probed by a comparison to the results in Sec.~\ref{sec:2}. 
Second, the hard-wall model may help us develop more analytic intuition because its eigenfunctions are basically trigonometric functions at high energy levels regardless of the potential hill inside the hard-wall potential. We will call the models in Sec.~\ref{sec:2} the ``soft-wall'' model to compare with the ``hard-wall'' model. 

\begin{figure}
\centering
     {\includegraphics[width=9cm]{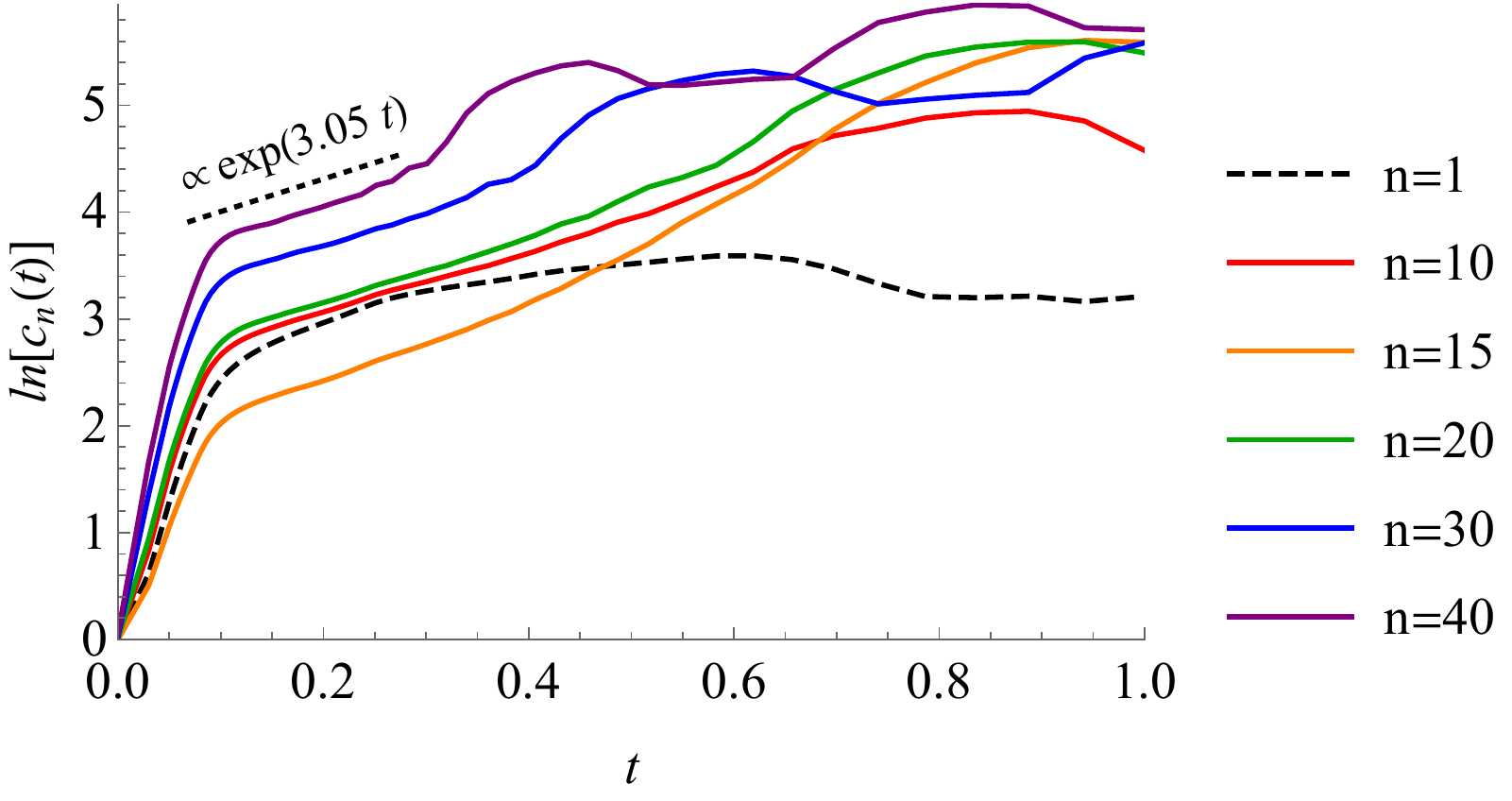} } 
 \caption{Time evolution of the microcanonical OTOCs for the model in  Fig.~\ref{pothhha} ($\lambda =2\sqrt{5}, g = 1/10$). The dotted curves do not have ranges of exponential growth.}\label{MC123}
\end{figure}

As a concrete example, we deal with the potential with $\lambda =2\sqrt{5}$ and $g = 1/10$, shown in Fig.~\ref{pothhha}. By the same procedures as Sec.~\ref{sec:2}, we compute the microcanonical OTOCs, some of which are shown in Fig.~\ref{MC123}. 
Let us compare the features with those of Sec.~\ref{sec:2}. 
There are two common features: 
i) The level close to the top of the hill has the steepest slope, {\it i.e.} the largest (microcanonical) Lyapunov exponent. In Fig.~\ref{MC123} this steepest slope  corresponds to $n=15$ (orange). ii) For small $n$ there is no exponential growth of the microcanonical OTOCs.  In Fig.~\ref{MC123} it corresponds to $n=1$ (dashed black). 
While we have these common features which are physically reasonable, 
there is a big difference from the soft-wall case in Fig.~\ref{fig:microcanonical}. As the energy level $n$ increases above the height of the potential hill, the microcanonical OTOCs still show the exponential growth,  while in Fig.~\ref{fig:microcanonical} in Sec.~\ref{sec:2} they are suppressed. The time range of the exponential growth decrease as $n$ increases. 

\begin{figure}
\centering
     {\includegraphics[width=9cm]{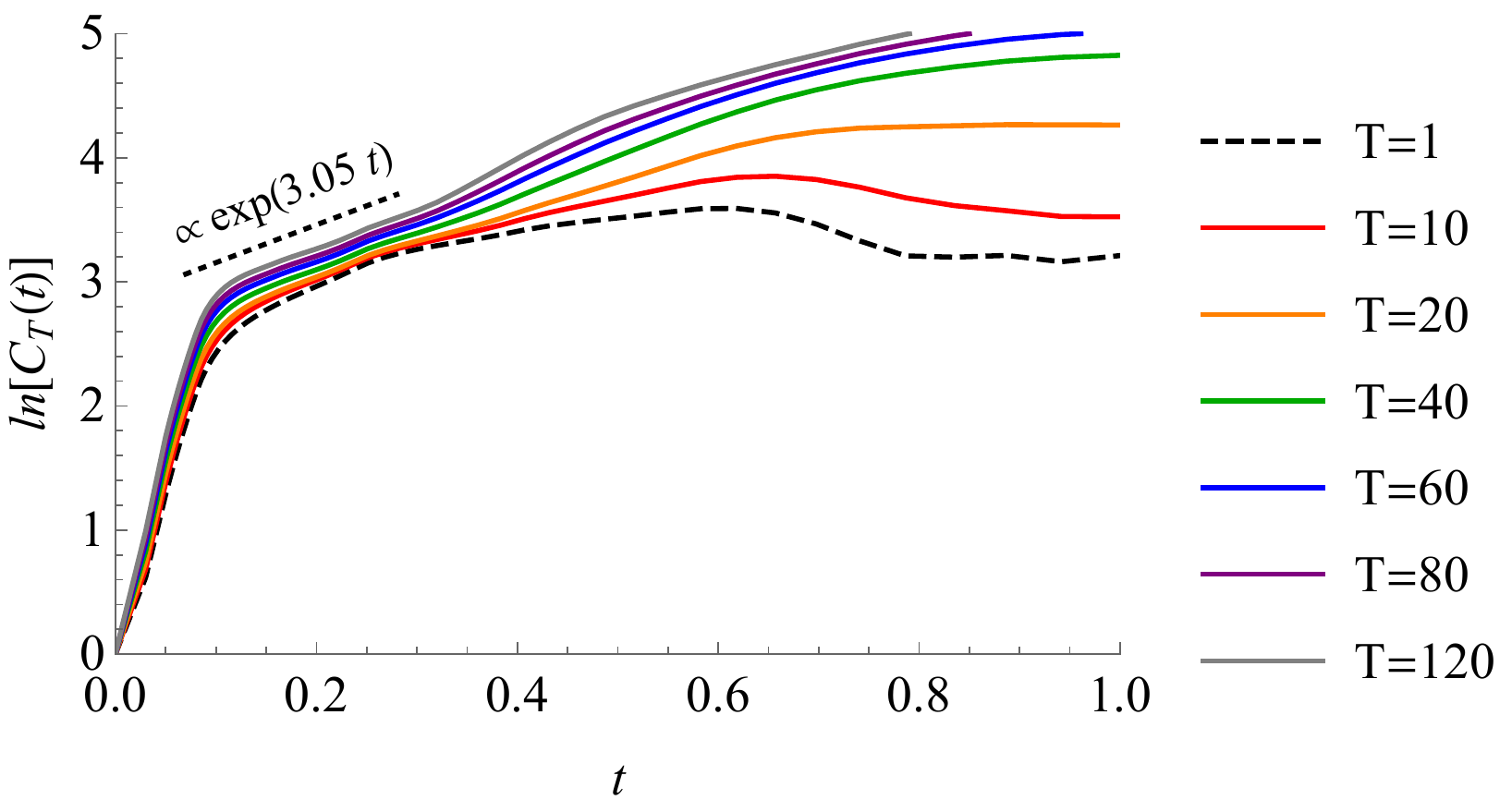} }
 \caption{Time evolution of the thermal OTOCs for the model in  Fig.~\ref{pothhha} ($\lambda =2\sqrt{5}, g = 1/10$). The dotted curves do not have ranges of exponential growth. }\label{T123}
\end{figure}

\begin{figure}
\centering
     \subfigure[Model in  Fig.~\ref{pothhha} ($\lambda =2\sqrt{5}, g = 1/10)$. ]
      {\includegraphics[width=6cm]{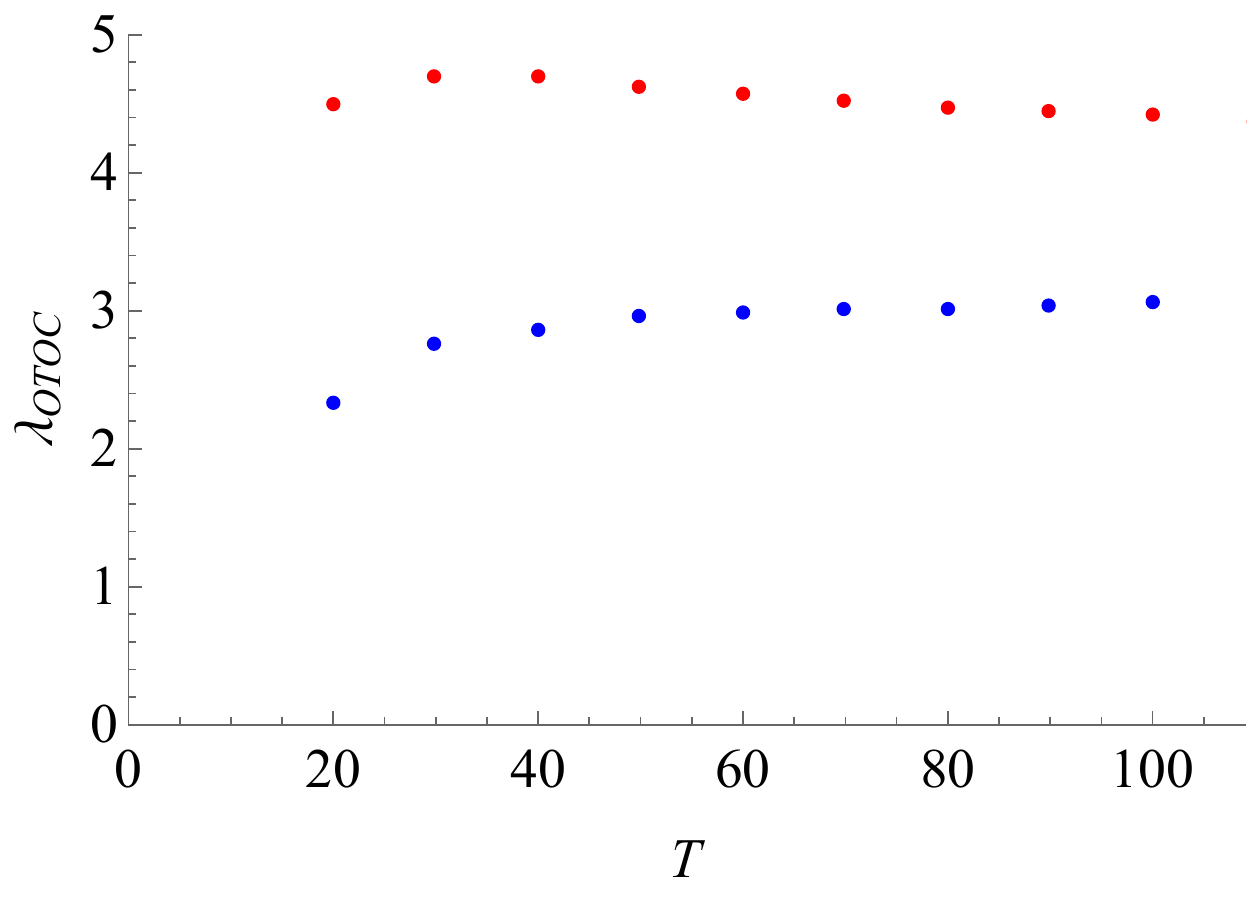} \label{QL123a}} \ \ \ 
          \subfigure[Model in  Fig.~\ref{pothhhb} ($\lambda =2, g = 1/50) $]
     {\includegraphics[width=6cm]{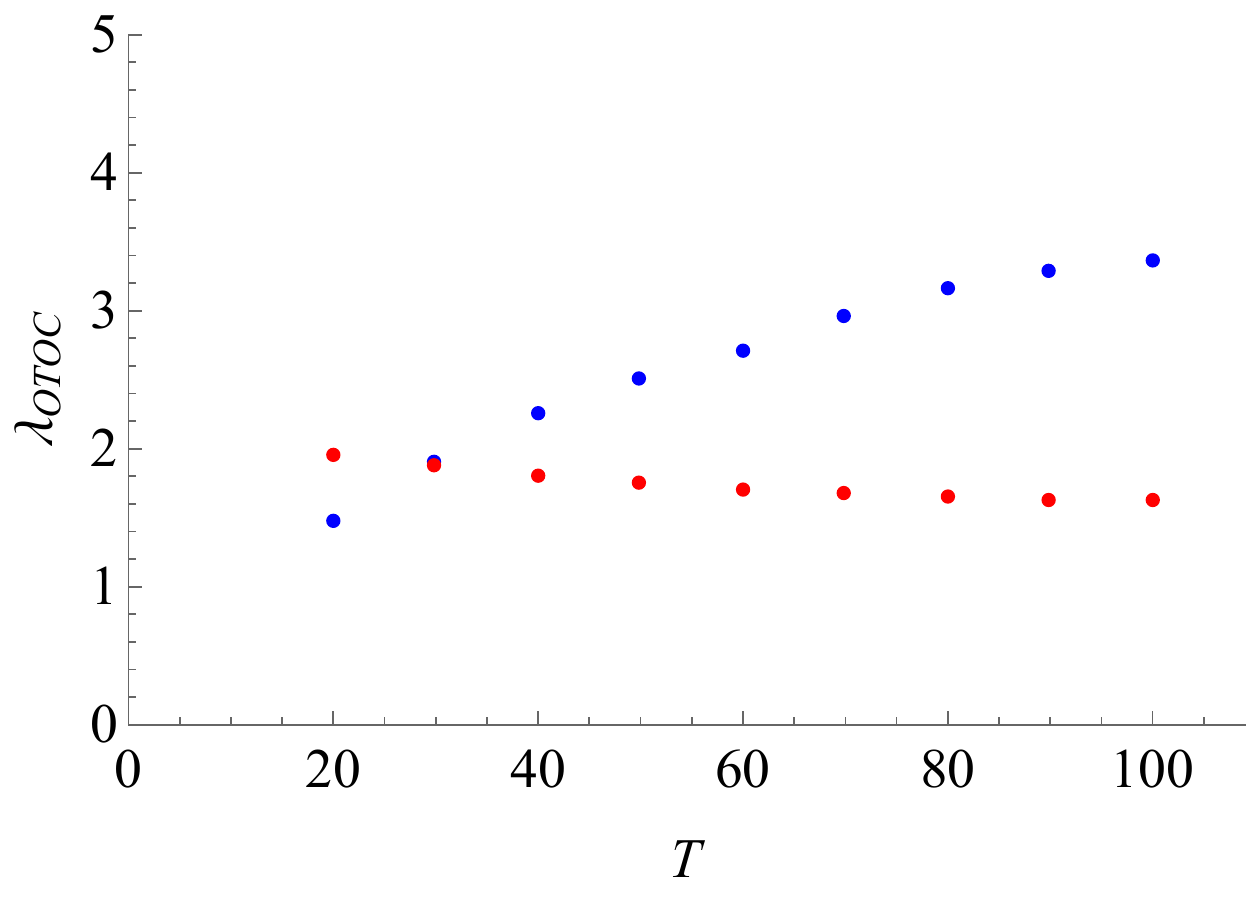} \label{QL123b}}
 \caption{Quantum Lyapunov exponent vs temperature. Red dots: soft-wall, Blue dots: hard-wall.}\label{QL123}
\end{figure}

By using the microcanonical OTOCs, we compute the thermal OTOCs, some of which at given values of the temperature are shown in Fig.~\ref{T123}. At low temperature, there is no exponential growth: see the dashed curve for $T=1$ case, for example. By reading off the slopes of the linear part of the curves in Fig.~\ref{T123} we make a plot of the quantum Lyapunov exponents at several values of the temperature. See the blue dots in  Fig.~\ref{QL123a}. In the infinite temperature limit, the quantum Lyapunov exponent saturates to $3.05$ approximately. For comparison, in Fig.~\ref{QL123a} the results of the soft-wall case are displayed as red dots. We find that the Lyapunov exponent asymptotes to a nonzero constant, and the value is ${\cal O}(\lambda_\text{saddle})$. 
These are common to what have been found in the previous section, and we find the universality.

By doing the same analysis for the hard-wall model with $\lambda =2$ and $g = 1/50$ shown in Fig.~\ref{pothhhb}, we obtain the blue dots in Fig.~\ref{QL123b}. The red dots are for the soft-wall case in Sec.~\ref{sec:2}. 
In this case, it is not clear if the quantum Lyapunov exponent saturates to a finite constant in the infinite temperature limit. The fitting with a $1/T$ expansion is a way to estimate the asymptotic value which turns out to be finite. To find a more reliable asymptotic behavior, we may need to investigate the higher temperature regime with more accuracy, against the numerical difficulties about the computational cost.

As seen in Fig.~\ref{QL123}, 
contrary to the soft-wall case in Sec.~\ref{sec:2}, 
the quantum Lyapunov exponents in hard-walls increase as temperature increases,
$d\lambda_{\rm OTOC}/dT\geq0$. This can be understood by the fact the microcanonical OTOCs are not suppressed as $n$ increases as shown in Fig.~\ref{MC123}. Furthermore, it  asymptotes to a function with a constant exponent, say $\bar{c}(t)$. 
In the infinite temperature limit, 
\begin{equation}
   	C_T(t) = \frac{1}{Z}\sum_n e^{-\beta E_n}c_n(t) \sim \bar{c}(t) \, .
\end{equation}
Thus, the quantum Lyapunov exponent of the thermal OTOC is equal to the microcannonical Lyapunov exponent at the large $n$ limit. Therefore, the exponential grwoth of the thermal OTOC is the accumulation effect of the microcannonical Lyapunov exponents of the {\it higher} modes rather than the strong effect of the microcannonical Lyapunov exponents of the {\it intermediate} levels near the saddle point.
For example, in Fig.~\ref{MC123}, we find that $c_n(t) \to \exp [3.05t]$ at large $n$, whose exponent is equal to the quantum Lyapunov exponent in   
Fig.~\ref{T123}. 
This feature is in strong contrast to that in the soft-wall models.
However, for both the hard-wall and the soft-wall cases the underlying physics comes from the saddle point. 
In particular, for the hard-wall case, it seems that the effect of the unstable maximum propagates among energy levels quite effectively and spreads to the whole system. 
This good efficiency may come from the commensurability of the energy levels and the simple trigonometric wave functions. So indeed the boundary walls of the potential affects the delicate behavior of the thermal OTOCs.

In spite of the difference in the high temperature behavior in the soft-wall and hard-wall models, the Lyapunove exponent asymptotes to finite values in the infinite temperature limit. The values slightly violate the semiclassical inequality \eqref{boundsaddle}, and again, this could be due to the quantum nature of the system.

\section{Lyapunov bound for quantum mechanics in one dimension}
\label{sec:4}

As described in the introduction, large $N$ quantum mechanical models may admit 
an effective description with just a single degree of freedom, and in such a case the chaos bound 
\eqref{MSSbound}
is expected also for a quantum mechanical model with just a single degree of freedom. Since such a quantum mechanics never have chaos, the only possibility is to have the unstable maximum in the potential to generate a nonzero Lyapunov exponent, in the manner described in Sec.~\ref{sec:2} and Sec.~\ref{sec:3} of this paper. With this motivation, we shall look for a mechanism of why \eqref{MSSbound} can work even in one-dimensional quantum mechanics. 

In fact, the results of Sec.\ref{sec:2} and Sec.~\ref{sec:3} show that all Lyapunov exponents measured by the thermal OTOCs satisfy \eqref{MSSbound}. The bound
\eqref{MSSbound} would have been violated if the exponential growth is seen
at the value of temperature below $\lambda_\text{saddle}/2\pi$, but this
value is too low for having the exponential growth, as observed in Fig.~\ref{fig:thermal_IHO}. This suggests that there may exist some quantum mechanism to prohibit going to lower temperature to violate the bound \eqref{MSSbound}.

In this section, we provide an intuitive explanation of the chaos bound \eqref{MSSbound} 
for generic quantum mechanics in one dimension. What we assume is 
that the exponential growth of the thermal OTOC, with the Lyapunov exponent $2\lambda$, is caused by 
a potential hill of the form of an inverted harmonic oscillator, whose classical Lyapunov exponent is $\lambda$.
Under this assumption, with generic quantum mechanical arguments, we can derive the bound \eqref{bound} for the Lyapunov exponent:
\begin{align}
\lambda \lesssim c\,  T \, , \quad c \simeq {\cal O}(1) \, .
\end{align} 

The principles which we use for our derivation of \eqref{bound} are the following natural facts which any quantum mechanical system is subject to.
For any quantum wave function of an energy eigenstate with energy $E$ to probe the local maximum, the following two conditions apply.
\begin{itemize}
\item {\bf Potential height condition.}
The energy $E$ of the quantum wave function can probe the local maximum only when the energy $E$
is larger than the height of the potential $V_\text{top}$, 
\begin{align}
E \gtrsim V_\text{top} \, .
\label{cond1}
\end{align} 
\item {\bf Quantum resolution condition.}
The quantum wave function can discriminate the local maximum only when the effective width $\Delta x$ of the 
hill-shaped potential 
is smaller than a half of the wave length of the wave function. The wave length $l$ of a plane wave 
and its energy $E$ are related as $E = \frac{(2\pi)^2}{l^2}$. So, the quantum resolution condition is
\begin{align}
E > \frac{\pi^2}{(\Delta x)^2} \, .
\label{cond2}
\end{align} 
\end{itemize}
Since the thermal OTOC is a summation of microcanonical OTOCs with the thermal weight $\exp[-E/T]$,
a necessary
condition for the thermal OTOC at temperature $T$ to probe the local maximum is, according to 
\eqref{cond1} and \eqref{cond2},
\begin{align}
T \gtrsim \text{max} \left\{ V_\text{top}, \, \frac{\pi^2}{(\Delta x)^2} \right\} \, .
\label{boundT}
\end{align} 
We evaluate the right hand side of this inequality to derive \eqref{bound}.

\begin{figure}[t]
\centering
	\includegraphics[width=130mm]{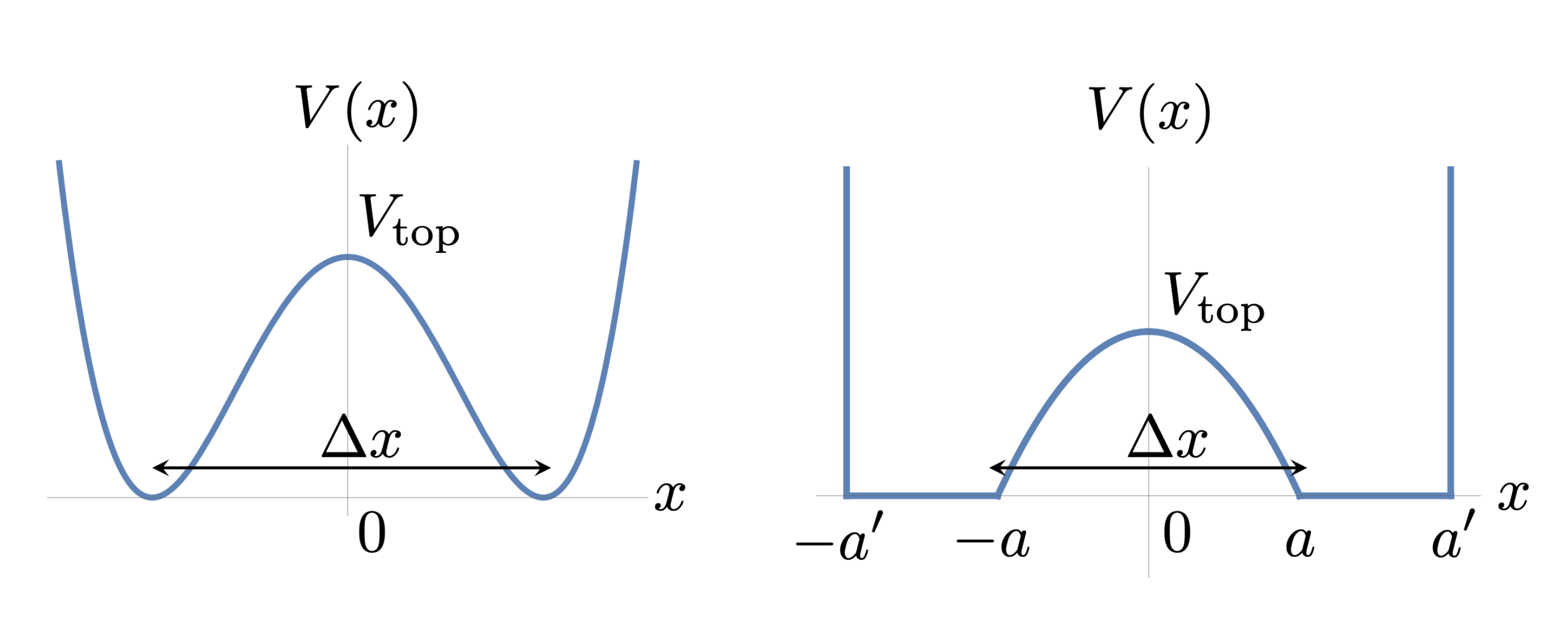}
	\caption{A schematic picture of the potentials we use for evaluating the bound for the Lyapunov coefficient.
	Left: the potential \eqref{Vquart}. Right: the potential \eqref{Vhardd}.}
	\label{fig:pot}
\end{figure}

To illustrate the generic statement, let us evaluate the right hand side of \eqref{boundT} 
with a concrete potential as the first example:
\begin{align}
V(x) = -\frac14 \lambda^2 x^2 + g x^4 + \frac{\lambda^4}{64g} \, ,
\label{Vquart}
\end{align} 
with $g>0$. See the left figure of Fig.~\ref{fig:pot}.
The last term is included so that the bottom of the potential is at $V=0$. 
The total Hamiltonian is $H = p^2 + V(x)$. This potential includes our analysis in Sec.~\ref{sec:2}
for some chosen values of $\lambda$ and $g$. The potential has a local maximum $x=0$, at which the classical 
Lyapunov exponent is $\lambda$.
In this case we find the height of the potential as
\begin{align}
V_\text{top} = \frac{\lambda^4}{64g}\, .
\end{align} 
The natural choice for the effective width of the potential is the distance between the two minima of the potential,
\begin{align}
\Delta x = \frac{\lambda}{\sqrt{2g}} \, .
\end{align} 
Using these, the inequality \eqref{boundT} is written as
\begin{align}
T > \text{max} \left\{ \frac{\lambda^4}{64g}, \frac{2\pi^2g}{\lambda^2} \right\} \, .
\end{align} 
Our goal is to find the most effective way to saturate this bound. Hence we change the potential while fixing 
$\lambda$ to find the minimum value of the temperature $T$. This is achieved by varying $g$ in the right hand side,
and the result is
\begin{align}
T_\text{min} = \frac{\sqrt{2}\pi}{8} \lambda \, 
\label{Tminl}
\end{align} 
with the optimized potential parameter 
\begin{align}
g= \frac{\lambda^3}{8\sqrt{2}\pi} \, . 
\label{optg}
\end{align}
This equation \eqref{Tminl} is equivalent to the bound
\begin{align}
\lambda < \frac{4\sqrt{2}}{\pi} T
\label{boundquart}
\end{align} 
which is nothing but \eqref{bound} which we wanted to show\footnote{The coefficient $4 \sqrt{2}/\pi$ is 
${\cal O}(1)$ and is less than $2\pi$.}.

\begin{figure}[t]
\centering
	\includegraphics[width=70mm]{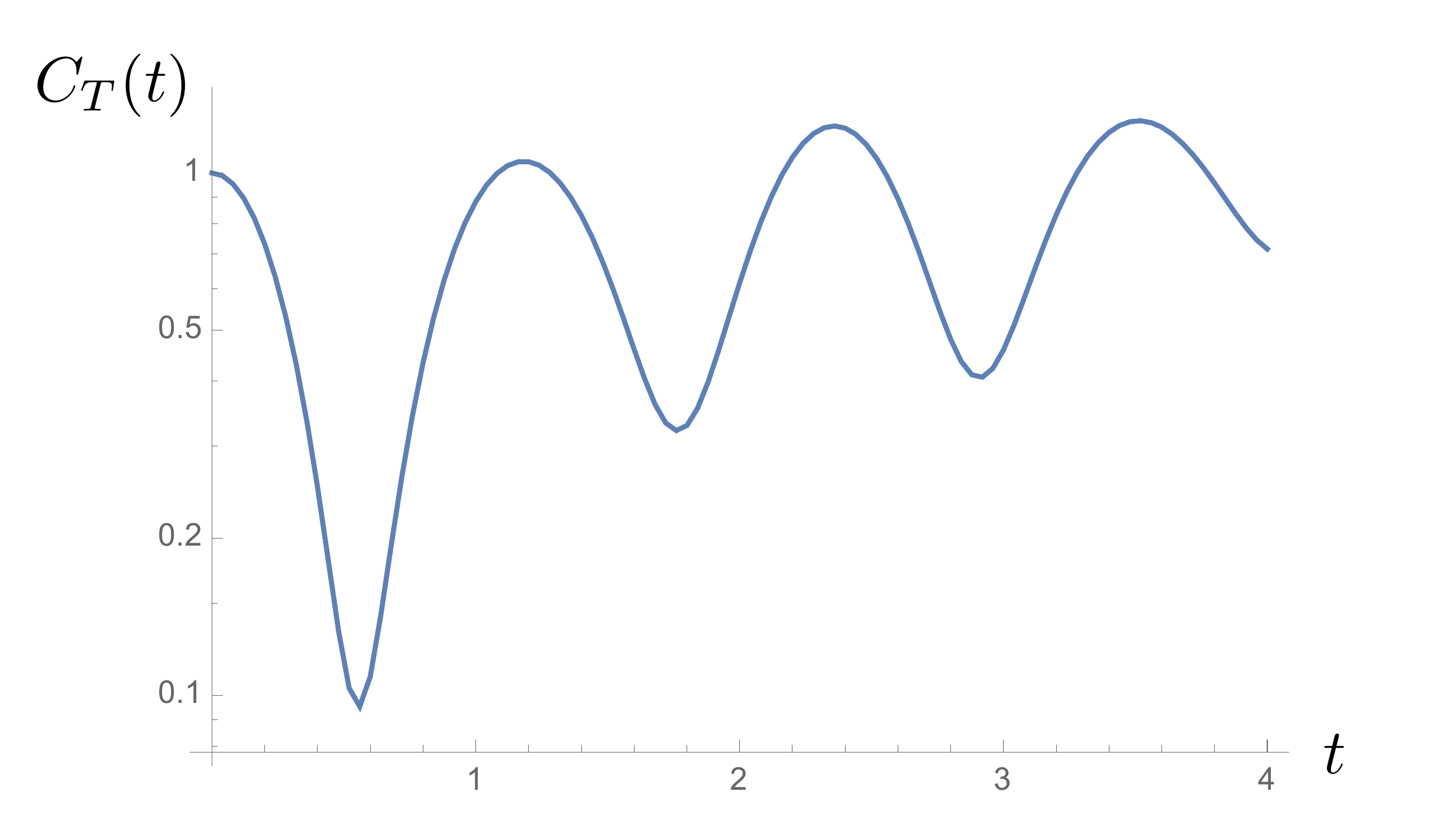}
	\caption{The time evolution of the thermal OTOC $C_T(t)$ of the system \eqref{Vquart}, with the optimized coupling $g$ \eqref{optg}, at the
	temperature value $T$ saturating \eqref{boundquart}. For the numerical calculation we chose $\lambda=2$. Obviously there is no exponential growth seen, and the minimal value of the temperature is too low to detect the local maximum effectively. 
	}
	\label{fig:boundT}
\end{figure}

It should be noted that this bound is the necessary condition, and for a given $\lambda$ the minimal value of
the temperature to observe the exponential growth of the thermal OTOC would be higher than the value 
saturating the inequality \eqref{boundquart}. To see this concretely, we numerically calculate the thermal OTOC
of the system \eqref{Vquart} with $\lambda=2$ and the value of $g$ tuned to satisfy \eqref{optg}. At the
temperature value saturating the inequality \eqref{boundquart}, the thermal OTOC is plotted in Fig.~\ref{fig:boundT}.
The OTOC does not show any exponential growth at this value of the temperature. Therefore, we are just looking at
necessary conditions for the exponential growth to be seen in the thermal OTOCs.

The inequality \eqref{bound} can be shown in a more general setup of the potential. Consider the potential
\begin{align}
V(x) = \left\{
\begin{array}{lll}
-\frac14 \lambda^2x^2 + \frac14 \lambda^2a^2 &\quad & (|x| \leq a)\\
0 &\quad &  (a \leq |x| \leq a') \\
\infty & \quad & (a' < |x|)
\end{array}
\right.
\label{Vhardd}
\end{align} 
See the right panel of Fig.~\ref{fig:pot}.
There exists a potential hill whose local maximum is at $x=0$.
The classical Lyapunov exponent at $x=0$ is taken to be $\lambda$, as in the previous case. 
The hard walls are located at $|x|=a'$. 
The model is similar to the one used in Sec.~\ref{sec:3}, and now we allow
arbitrary location of the hard walls. In fact, in the following discussion, the
potential shape in the region $|x|>a$ does not matter.

Since the bottom of the potential is $V=0$, we find
\begin{align}
V_\text{top} = \frac14 \lambda^2 a^2 \, .
\end{align} 
The effective width of the potential hill is obviously
\begin{align}
\Delta x = 2a \, .
\end{align} 
Then the bound for the temperature of the thermal OTOC \eqref{boundT} is\footnote{
In the right hand side of \eqref{T>max}, the quantity $\frac{\pi^2}{4a^2}$
happens to be equal to the zero-point energy for the case of a single-well potential with the size $2a$.
In this sense, one may think that 
the quantum resolution condition may be rephrased as the condition that the temperature is
larger than the order of the ground state energy. 
But this condition can always be achieved by simply taking 
$a' \to \infty$, while the quantum resolution condition in fact forbids this limit.
}
\begin{align}
T > \text{max} \left\{ \frac14\lambda^2a^2, \frac{\pi^2}{4a^2} \right\} \, .
\label{T>max}
\end{align} 
The right hand side is minimized when $a= \sqrt{\pi/\lambda}$, at which we find
\begin{align}
\lambda < \frac{4}{\pi} T \, .
\end{align} 
This is again the inequality \eqref{bound} with the ${\cal O}(1)$ numerical coefficient\footnote{The coefficient $4/\pi$ is less than $2\pi$.}.
Note that this argument does not depend on $a'$. 
Thus we can generally expect that the argument above will not 
depend on the structure of the potential outside of the inverted harmonic oscillator part, and we have the generic bound \eqref{bound} for any bounded potential which includes the inverted harmonic potential.

In this section, we have provided a derivation of \eqref{bound} which is
of the same form as the chaos bound discovered in \cite{Maldacena:2015waa}.
The latter is the bound for chaotic large $N$ systems, while our bound \eqref{bound}
is for one-dimensional quantum mechanical systems which are classically non-chaotic.
Possible concrete relations between the two, if any along the direction described in the introduction, would be interesting.

\section{Summary and discussions}
\label{sec:5}

In this paper we have investigated Lyapunov exponents $\lambda_\text{OTOC}$ of the thermal OTOCs for one-dimensional quantum mechanical systems with an inverted harmonic oscillator potential. The system is non-chaotic, and the classical counterparts are general with a local maximum which can generate a local classical Lyapunov exponent $\lambda_\text{saddle}$. We have numerically evaluated $\lambda_\text{OTOC}(T)$ for various values of temperature. We have discovered that
at values of the temperature above a certain threshold the exponential growth is observed in the thermal OTOCs, and the measured $\lambda_\text{OTOC}(T)$ is of the same order as $\lambda_\text{saddle}$. As we extrapolate our numerical results to $T=\infty$, the Lyapunov exponent $\lambda_\text{OTOC}(T=\infty)$ is suggested to be non-vanishing. These features are shared in various quantum mechanical models and universal, as we studied in detail in Sec.~\ref{sec:3}.  
Our results of $\lambda_\text{OTOC}(T)$ for the Higgs-type potential case are summarized in Fig.~\ref{fig:Lyapunov_IHO} and Fig.~\ref{pot2b} in Sec.~\ref{sec:2}, and for the hard-wall potential case in Fig.~\ref{QL123} in Sec.~\ref{sec:3}. 

Our findings have shown that the thermal OTOC can grow exponentially in time generically in one-dimensional quantum mechanics which are regular (non-chaotic). The temperature dependence of the OTOCs confirms that the origin of the exponential growth is a classical Lyapunov exponent at the saddle (the local maximum) of the potential. Since this is shown in our generic one-dimensional systems, it is natural to expect that finite-dimensional quantum systems follow the same behavior. If we equate the exponential growth of the thermal OTOC with the information scrambling at finite temperature, we are led to the conclusion that the information scrambling can happen in non-chaotic quantum systems.

At low temperature the exponential growth cannot be numerically identified, while the Lyapunov exponent $\lambda_\text{OTOC}$, if observed, needs to be ${\cal O}(\lambda)$ which is fixed by the curvature of the potential at the unstable maximum. This suggests that there exists a bound concerning the Lyapunov exponent and the temperature, which is suggestive in view of the ``chaos bound'' \eqref{bound} \cite{Maldacena:2015waa}. 
In Sec.~\ref{sec:4} we have derived a bound \eqref{boundT}, $\lambda_\text{OTOC}(T) \lesssim c \, T$ with $c \simeq {\cal O}(1)$ for generic one-dimensional quantum systems. This bound is simple and quite similar to \eqref{bound}. The derivation is based on two facts which are satisfied generically in quantum mechanics in one dimension: first, the energy of the wave function to probe the local maximum needs to be higher than the potential energy of the maximum, and second, the wave length needs to be shorter than the scale of the potential hill. It is surprising that such a simple bound of the form \eqref{bound} and \eqref{boundT} holds for a wide class of quantum systems.

Several discussions on our results are in order. 
First, our $\lambda_\text{OTOC}$ evaluated at $T=\infty$ by a fitting does not satisfy the semiclassical inequality \eqref{boundsaddle}, as described in Sec.~\ref{sec:2} and Sec.~\ref{sec:3}. This could be due to the fact that our analyses are not semiclassical but fully quantum, and/or the fact that the Hilbert space of our system is infinite dimensional. Pinning down the reason would help us when we generalize the analyses to quantum field theories which have much bigger Hilbert spaces\footnote{See \cite{Stanford:2015owe} for an example of the evaluation of the thermal OTOC in a quantum field theory.}, in view of the holographic principle. Numerical investigation of the semiclassical limits of our system, and comparison to the general semiclassical analyses \cite{Jalabert:2018kvl}, may provide a path to a resolution. 

Next, in our bound \eqref{boundT}, the numerical coefficient $c$ is dependent on what kind of potential one chooses for the walls. A natural question is whether $c=2\pi$ or not, to compare \eqref{boundT} with \eqref{bound}. In fact, it is difficult to find the exact value of $c$ which can work for any system, because the principles we use for the derivation is difficult to be quantified: the wave length needs to be smaller compared to the length scale of the potential hill, but here, the ``length scale'' is ambiguous. Therefore, to make a precise statement with some explicit number $c$, we may need to introduce a measure of the detectability of the curvature of the potential by wave functions.

Finally, as described in the introduction, finding any possible relation between the chaos bound \eqref{bound} and our quantum mechanical bound \eqref{boundT} would be interesting. It may open up a bridge between large-$N$ quantum mechanics and few-body quantum mechanics, through information scrambling, chaos and holographic principle. We like to revisit the issues in the future.



\acknowledgments

K.~H.~and R.~W.~would like to thank Lea Ferreira dos Santos and Sa\'ul Pilatowsky
for valuable discussions which motivated the present work. K.~H.~would like to thank Takeshi Morita 
for valuable discussions.
K.~H.~was supported in part by JSPS KAKENHI Grant No.~JP17H06462.
K.-Y.~K.~and K.-B.~H.~were supported by Basic Science Research Program through the National Research Foundation of Korea(NRF) funded by the Ministry of Science, ICT $\&$ Future Planning(NRF- 2017R1A2B4004810) and GIST Research Institute(GRI) grant funded by the GIST in 2020.


\begin{figure}[]
	\centering
		\includegraphics[width=75mm]{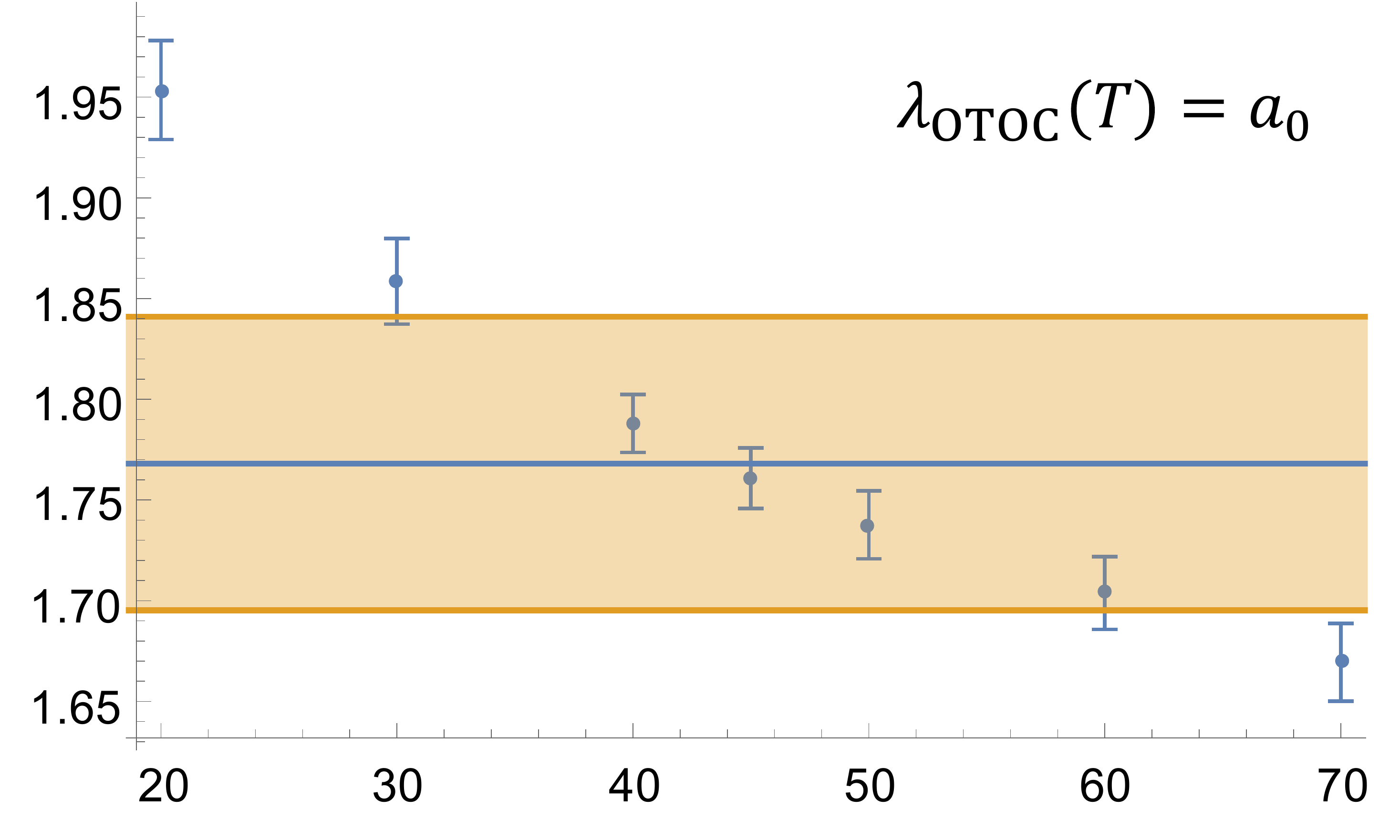}
		\includegraphics[width=75mm]{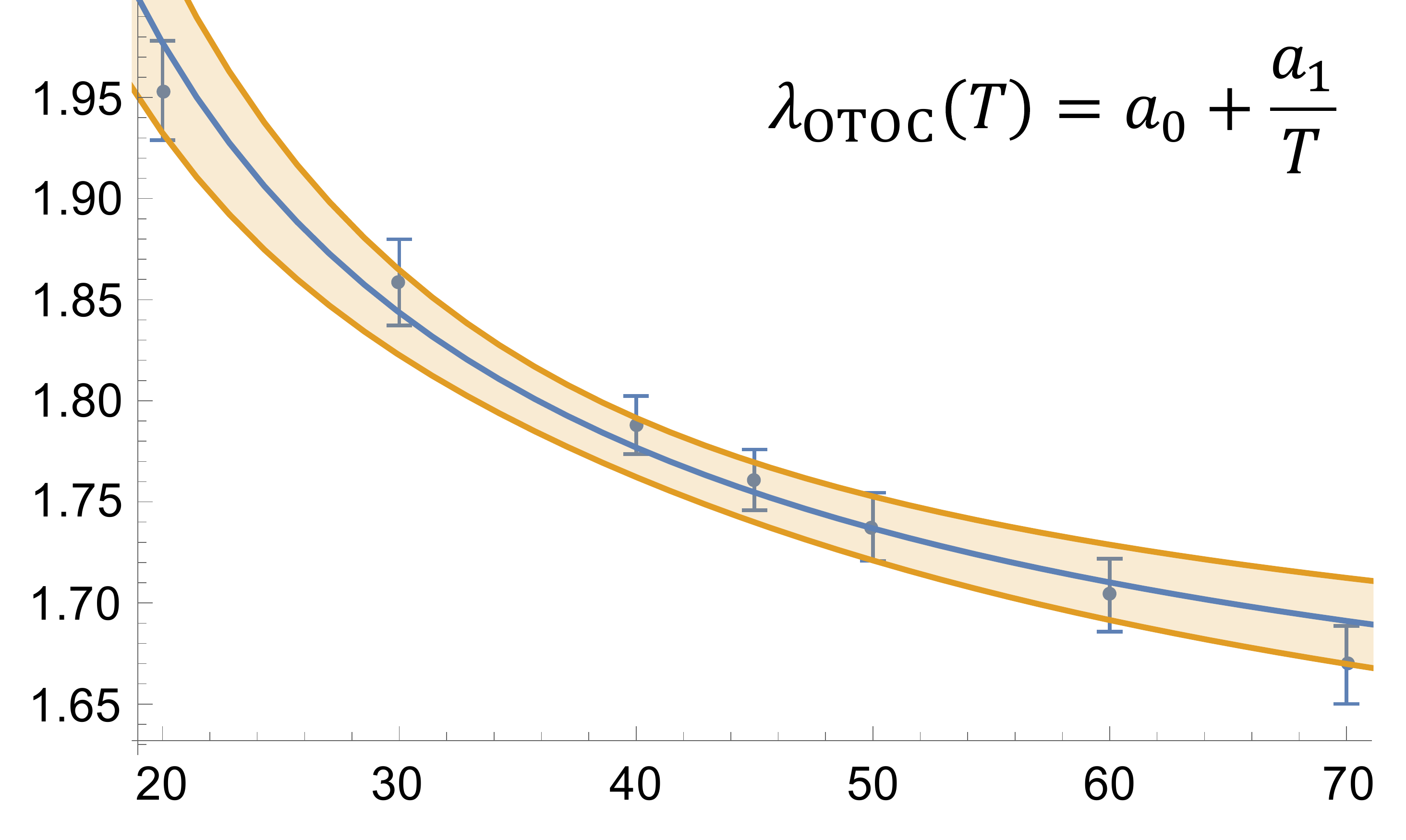}\\
		\includegraphics[width=75mm]{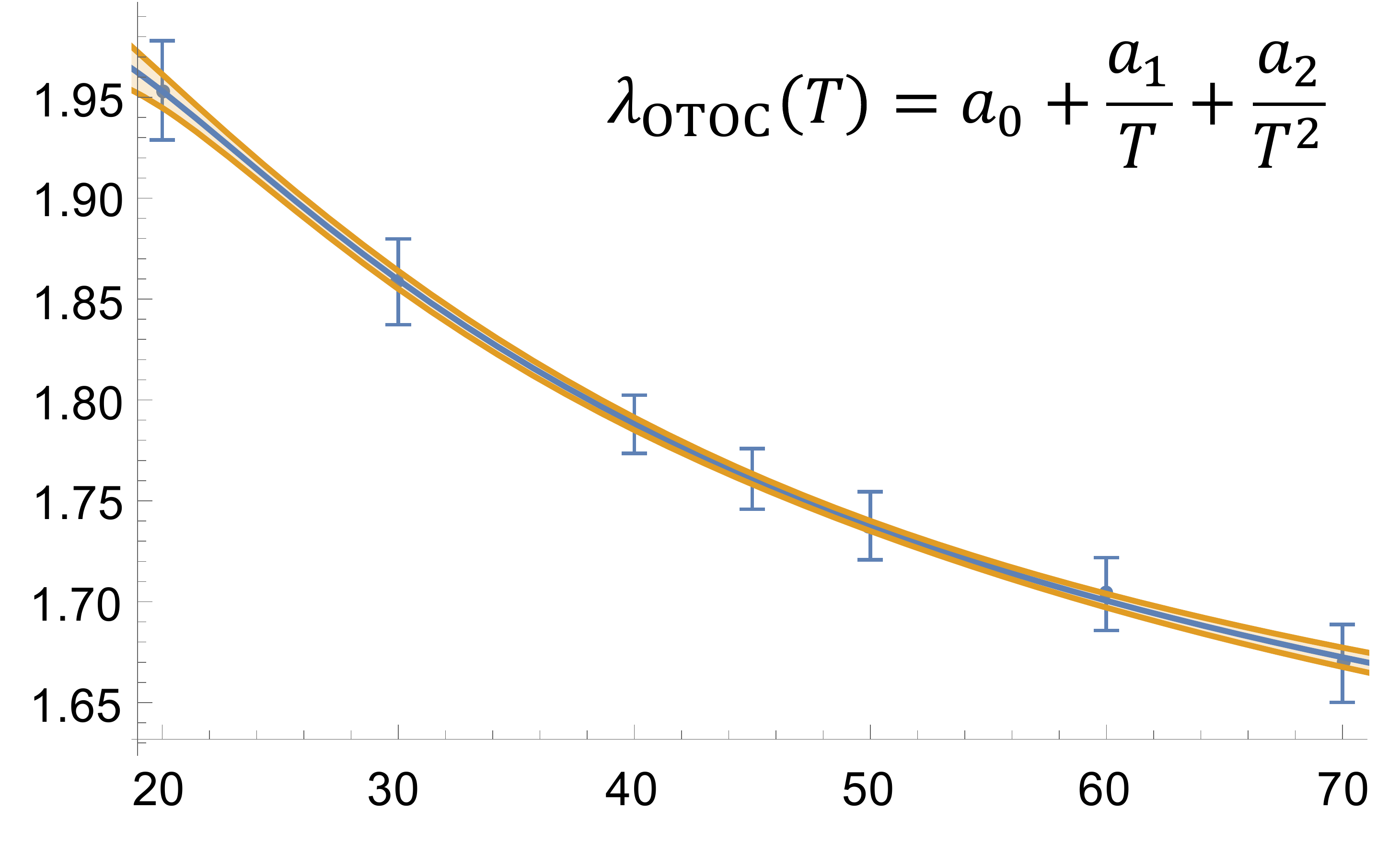}
		\includegraphics[width=75mm]{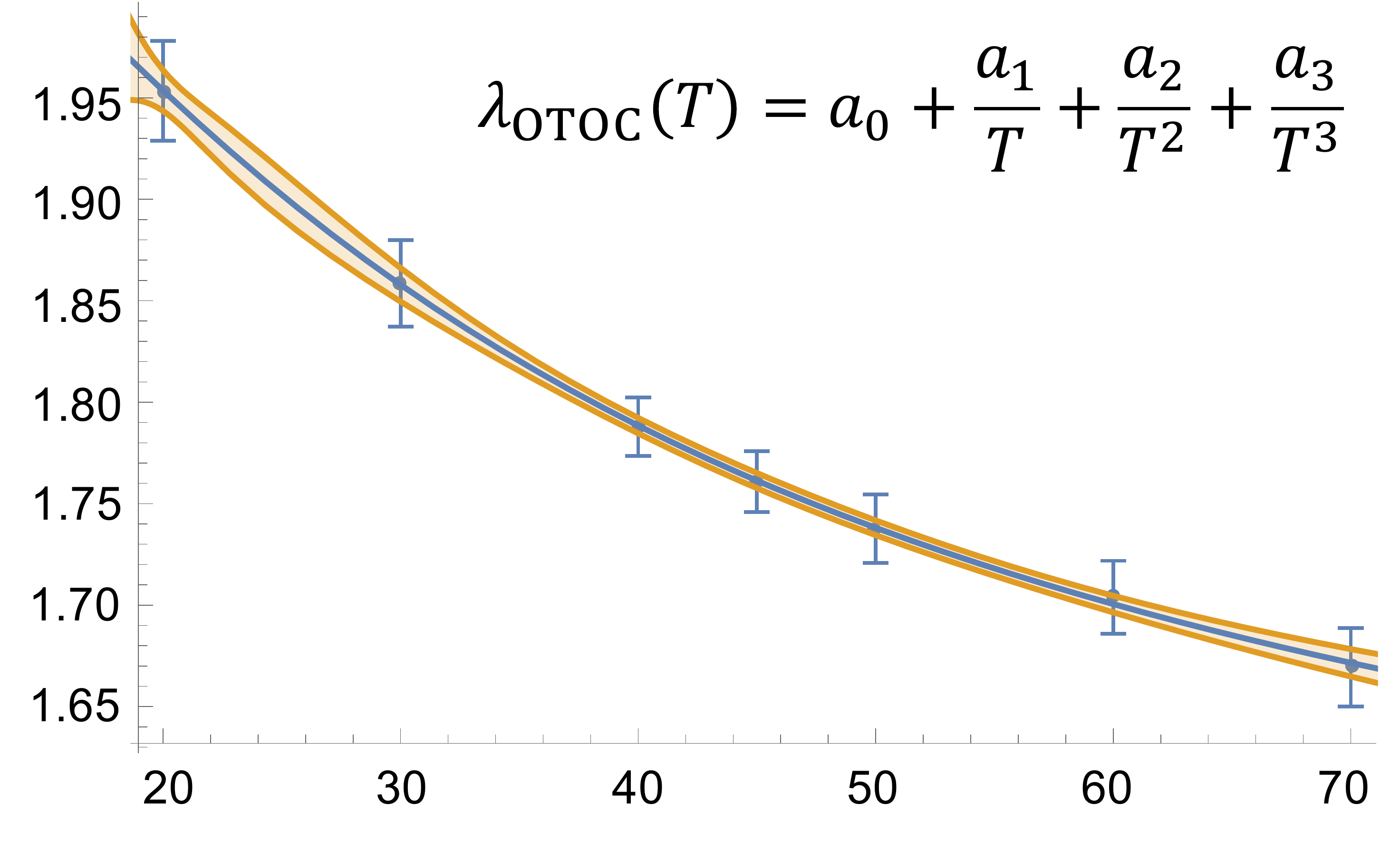}
		\caption{Blue curves (lines): the fitting function. Yellow region: 95\% confidence interval for the fitting.}
		\label{fig:the temperature dependence}
\end{figure}

\appendix

\section{Error analysis of high-temperature fitting of Lyapunov exponents}
\label{App:A}

In Sec.~\ref{sec:2}, we have assumed that the Lyapunov exponent $\lambda_{\rm OTOC}(T)$ can be expanded as \eqref{eq:expansion}. 
For the fitting, we consider the following four possibilities: 
\begin{equation}
\label{eq:functional forms}
	\lambda_{\rm OTOC}(T) = a_0,\, ~  a_0+\frac{a_1}{T},\, ~ a_0+\frac{a_1}{T}+\frac{a_2}{T^2},\, ~ a_0+\frac{a_1}{T}+\frac{a_2}{T^2}+\frac{a_3}{T^3} \, .
\end{equation}
The results of the fitting with each of these functions are shown in Fig.\ref{fig:the temperature dependence}. The blue curve is the fitting function and the yellow region is its 95\% confidence interval.
From these plots, we conclude that the fitting with the form 
$a_0 + a_1/T $ is the most credible.
For the fitting to be reasonable, it must be the same degree of accuracy as the error bars of the data points. From this point of view, the last two fittings are obviously overtrained. On the other hand, in the second fitting $a_0 + a_1/T $, the accuracy of the fitting is the same order as that of the data points. 

It is interesting to note that the inclusion of the order $1/T^2$
can reproduce also the the exponents at lower temperature in Fig.~\ref{fig:Lyapunov_IHO}. This is suggestive to further explore the
whole structure of the temperature dependence of the Lyapunov exponent
$\lambda_\text{OTOC}$.


\section{Other operator orderings and the origin of the exponential growth}
\label{App:B}

The OTOC which we evaluate in this paper is the one with the commutator squared, \eqref{eq:thermal OTOC} and \eqref{eq:microcanonical OTOC}. More generally, as is found in literature, 
one can also consider the OTOC of the form
\begin{equation} \label{Eq:Ft}
  F(t) \equiv \langle x(t) p(0)x(t)p(0) \rangle \,, 
\end{equation}
without using the commutator. In the semiclassical analysis,
the expected general behavior is $F(t) \sim \text{const.} + \hbar \exp[\lambda t] + \cdots$, while
our commutator squared OTOC may result in $\langle [x(t), p(0)]^2 \rangle \sim \hbar^2 \exp[2 \lambda t] + \cdots$,
by a naive replacement of the Poisson bracket with the commutator (here note that the phase space volume suppression considered in \cite{Xu:2019lhc} is ignored). 
One might think that this difference could affect\footnote{
This viewpoint was brought to us by Takeshi Morita, and we would like to thank him.
} 
the comparison
of the OTOC Lyapunov exponent $\lambda_\text{OTOC}$ and the classical saddle Lyapunov exponent $\lambda_\text{saddle}$, by the factor of $2$, concerning the chaos bound \eqref{MSSbound}, because the
bound was derived through the form of $F(t)$.
However, let us recall that our commutator squared OTOC is supposed to be compared with Re$F(t)$, not $F(t)$ (see \eqref{eq:b4} or \eqref{CF}).  Because the leading term of Re$F(t)$  is $ \sim \hbar^2 \exp[2\lambda t]$, we may expect that Re$F(t)$ has the same Lyapunov exponent as our commutator squared OTOC without the factor 2 difference.

Upon this motivation, in this appendix, we present the evaluation of the OTOC of the form $\langle x(t) p(0) x(t) p(0)\rangle$. Let us start with the relation between $\langle x(t) p(0) x(t) p(0)\rangle$ and $\langle [x(t), p(0)]^2 \rangle$.  For the microcanonical OTOC \eqref{eq:microcanonical OTOC}, 
\begin{eqnarray}
   c_n &=&  - \langle n| [x(t), p(0)]^2 | n \rangle  \nonumber\\
   & =&  \bigg\{\langle n | x(t) p(0)^2 x(t) | n \rangle + \langle n| p(0) x(t)^2 p(0)  | n \rangle \bigg\} - 2 \mathrm{Re}\langle n| x(t) p(0) x(t) p(0) | n \rangle  \,,  \label{eq:b2}
\end{eqnarray}
and for the thermal OTOC \eqref{eq:thermal OTOC}
\begin{eqnarray}
   C_T &=&  - \langle  [x(t), p(0)]^2 \rangle 
   \nonumber \\
   & =&   \underbrace{\bigg\{\langle  x(t) p(0)^2 x(t) \rangle + \langle p(0) x(t)^2 p(0)  \rangle \bigg\}}_{\equiv G(t)} - 2 \mathrm{Re} F(t)\label{eq:b4}  \,, 
\end{eqnarray}
where \eqref{Eq:Ft} with the thermal average is used. 

Note that the time-dependence of Re$F(t)$ is not determined only by $C_T$ because of $G(t)$. The term, $G(t)$,  would have been $2$ if we started with unitary operators. i.e. for the Hermitian and unitary operators $V$ and $W$, the relation \eqref{eq:b4} yields
\begin{align}
C_T = -\langle [V(t), W(0)]^2 \rangle = 2 - 2 \, \text{Re}\,  F_{VW}(t) \,,
\label{CF}
\end{align}
where $F_{VW}(t) \equiv \langle V(t)W(0)V(t)W(0) \rangle$. In our case, we have chosen the non-unitary operators 
$x$ and $p$ to make the analogue to the classical deviation of the path in the $x$ space. 

In our case, although $G(t)$ is not constant,
it is still possible that the time-dependence of Re$F(t)$ is closely correlated with $C_T$ if $G(t)$ is effectively constant for a certain time range where  the Lyapunov exponent is defined. We check this possibility below, and it turns out that this is not the case. 
\begin{figure}
\centering
     \subfigure[ The components 2Re$F(t)$ and $G(t)$ in \eqref{eq:b4}.]
     {\includegraphics[width=6cm]{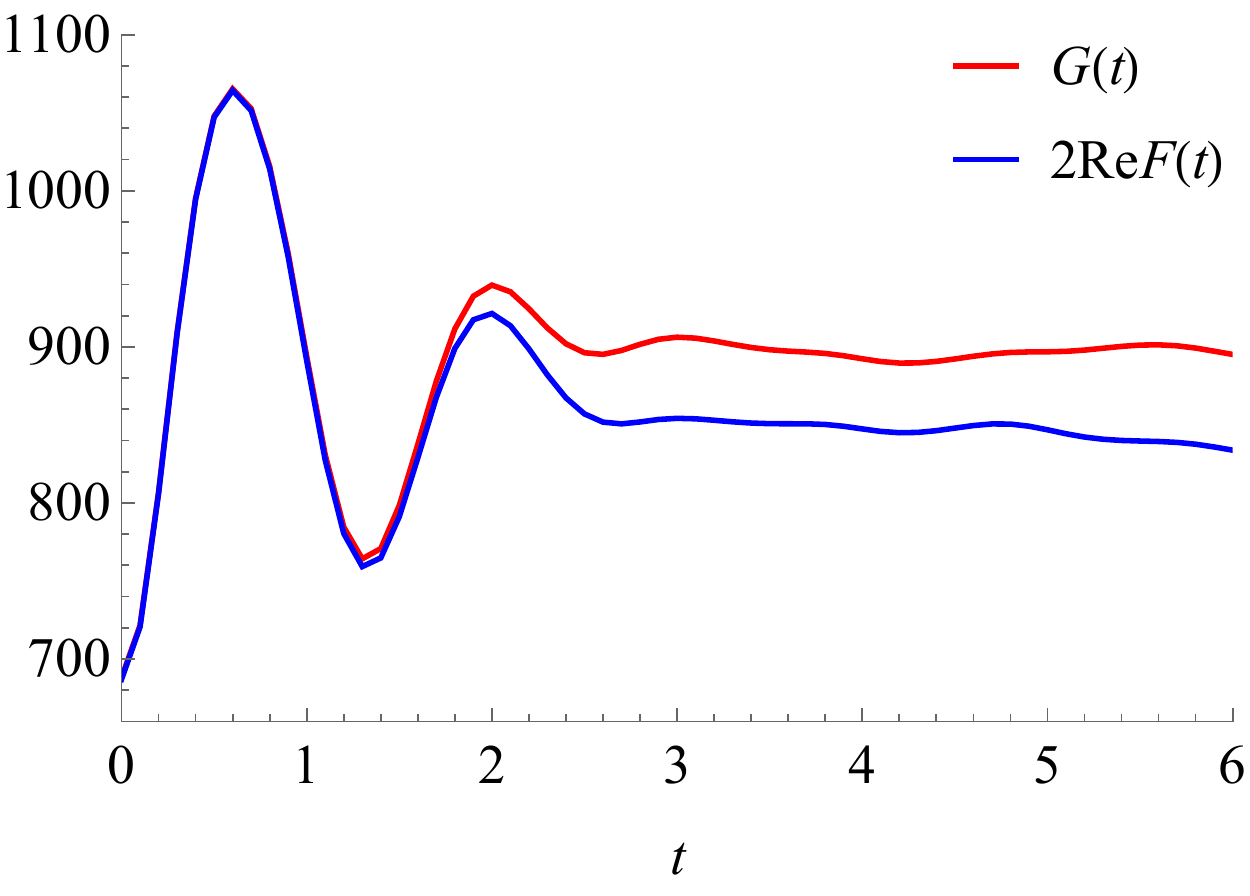} \label{XPXP_Thermal_1}} \ \ \  
     \subfigure[The total $C_T$ in \eqref{eq:b4}.]
     {\includegraphics[width=6cm]{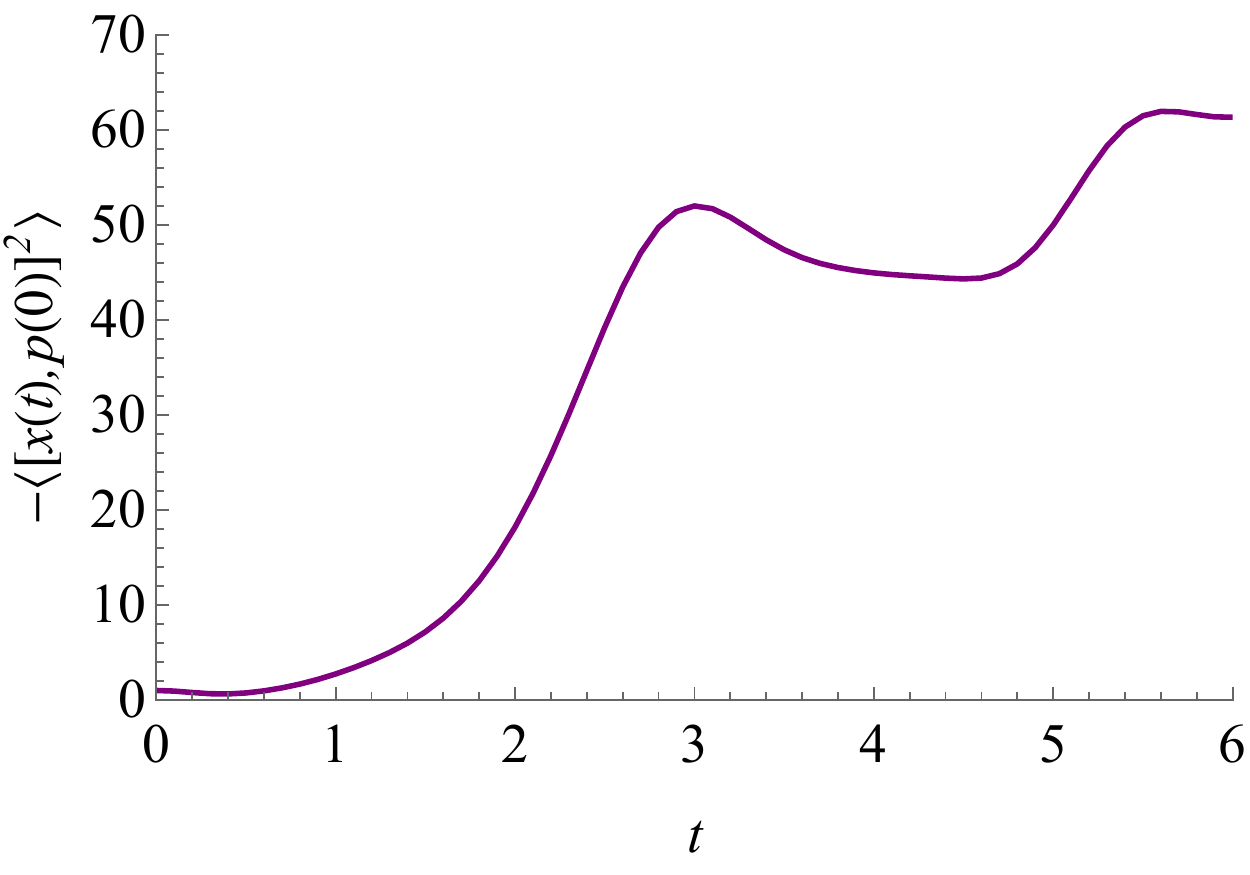} \label{XPXP_Thermal_2}}
     \caption{Time evolution of the thermal OTOC at $T=30$.}\label{XPXP_Thermal}
\end{figure}
For example, let us consider the model we studied in Fig.~\ref{fig:energy_eigenvalues}. Our thermal OTOC, $C_T$ in \eqref{eq:b4}, at $T=30$ is reproduced in 
Fig.~\ref{XPXP_Thermal_2} and its decomposition (the two terms in \eqref{eq:b4}) is shown in Fig.~\ref{XPXP_Thermal_1}. Here, the blue and red curve represent 2Re$F(t)$ and $G(t)$ respectively.
The exponential growth is observed in neither 2Re$F(t)$ nor $G(t)$. However, their difference (the red curve minus the blue curve in Fig.~\ref{XPXP_Thermal_1}) yields the exponential growth between $t \sim 1$ and $t \sim 3$ (Fig.\ref{XPXP_Thermal_2}).
Thus, the property of the exponential growth of $C_T$ is not shared by 2Re$F(t)$ in our model.

\begin{figure}
\centering
     {\includegraphics[width=4.5cm]{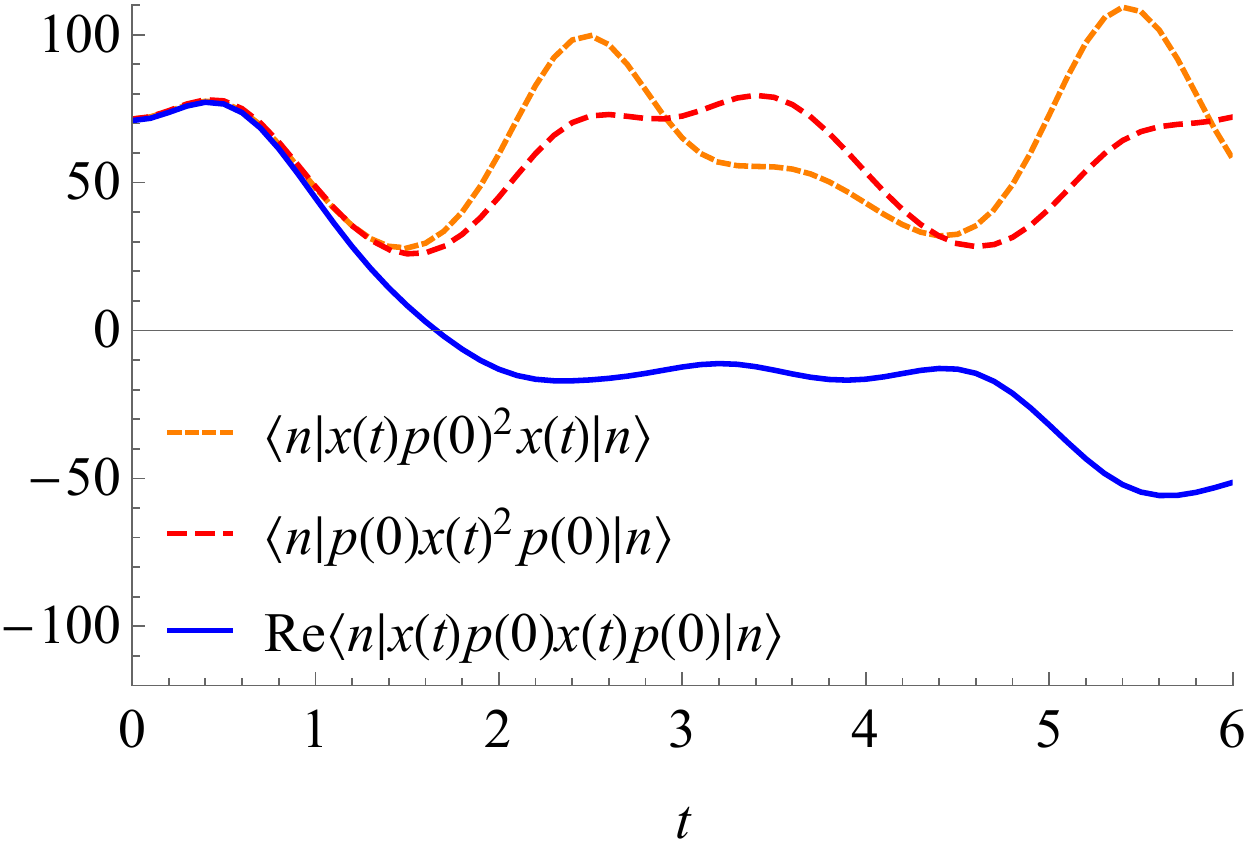} \label{XPXP_micro_1}} \ \ \
     {\includegraphics[width=4.5cm]{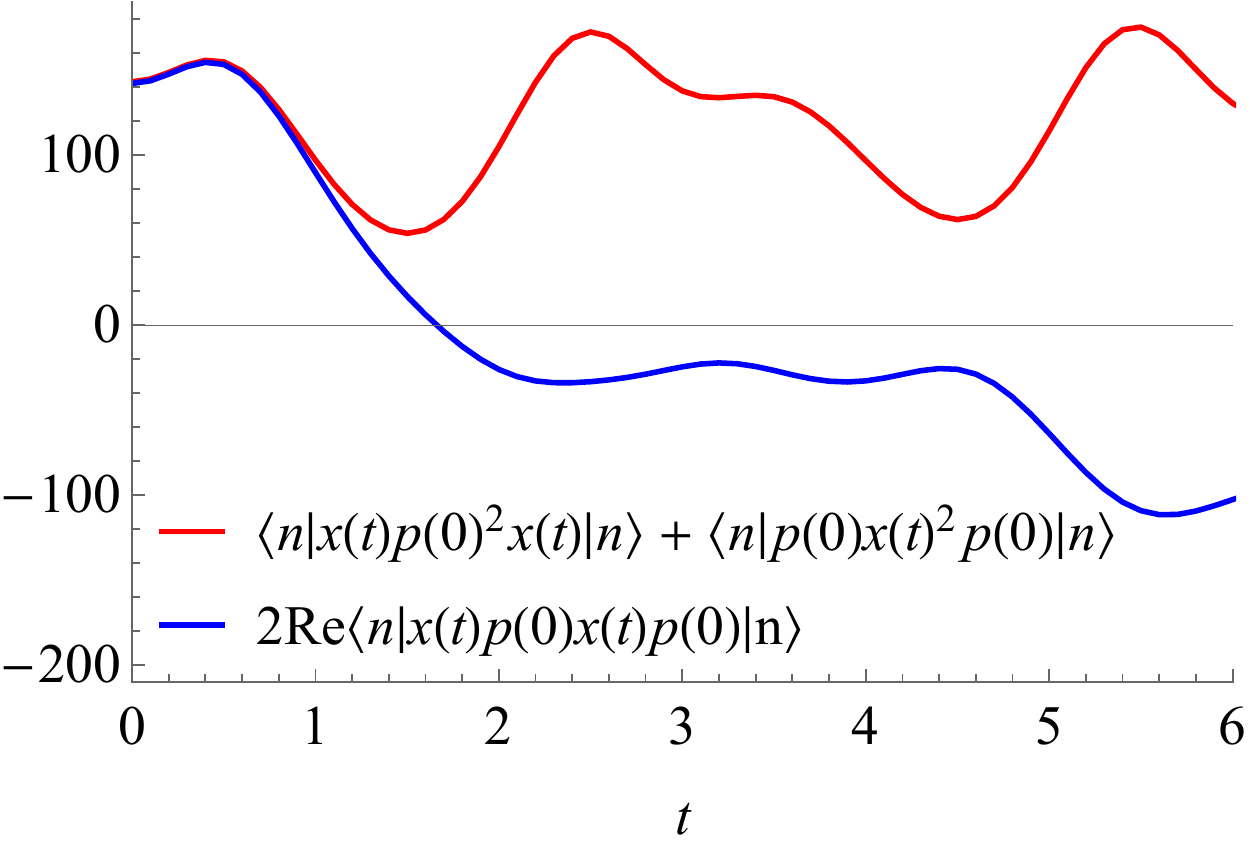} \label{XPXP_micro_2}}
     \ \ \
     {\includegraphics[width=4.5cm]{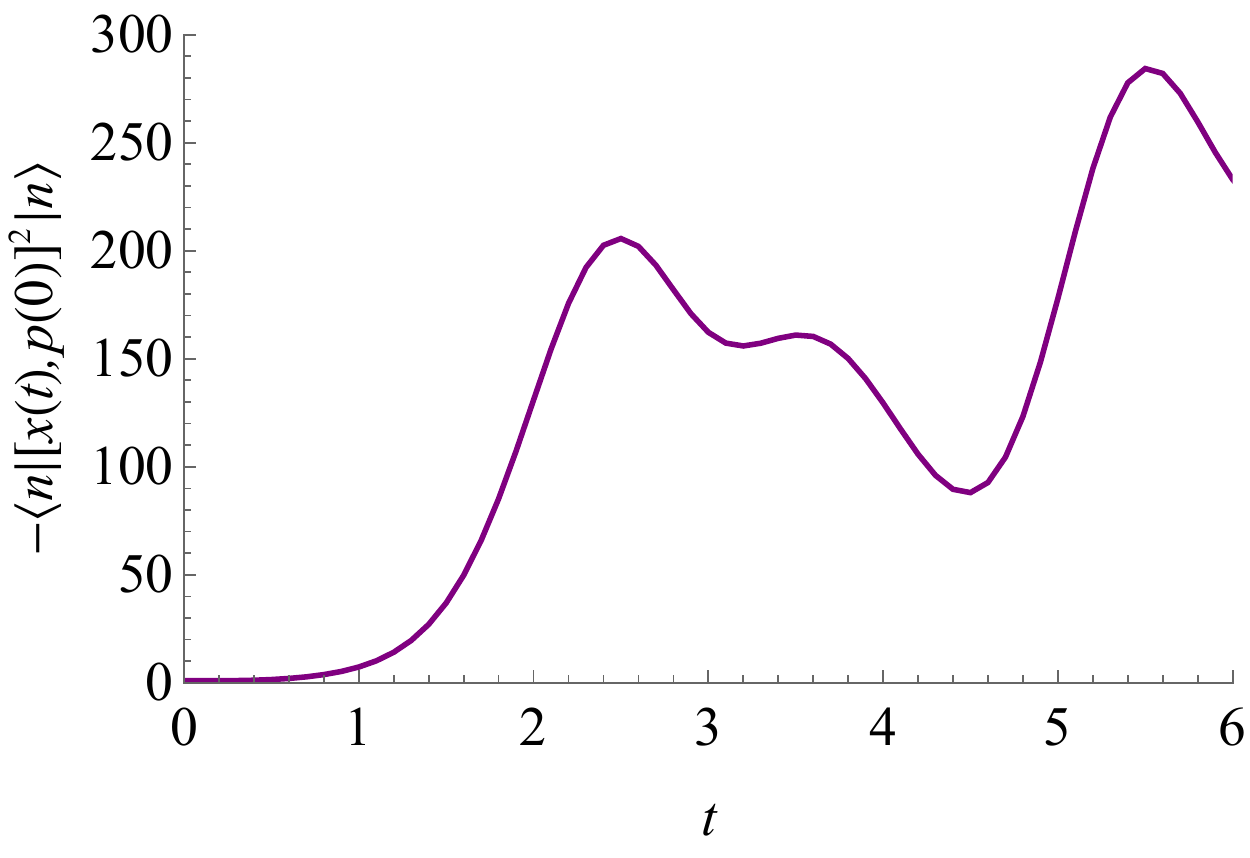} \label{XPXP_micro_3}}
 \caption{Time evolution of the microcanonical OTOC at  $n=11$.}\label{XPXP_micro}
\end{figure}

For completeness, we also show the microcanonical OTOC at $n=11$ in Fig.\ref{XPXP_micro}. The left figure represents three terms in \eqref{eq:b2}.
The middle and right one correspond to Fig.\ref{XPXP_Thermal_1} and Fig.\ref{XPXP_Thermal_2} respectively in the thermal OTOC case. 

One might argue that the decrease observed in Re$\langle n|x(t)p(0)x(t)p(0)|n\rangle$ (the blue curves in Fig.\ref{XPXP_micro}) could be exponential in the time range $0.5\lesssim t \lesssim 1.5$.
However, this time range turns out to be not long enough compared with the inverse of the corresponding exponent.
Thus, it is difficult to observe the exponential growth only from the last term in \eqref{eq:b2}.

Thus, here we conclude that the OTOC without using the commutator does not exhibit the exponential behavior, as opposed to our commutator squared OTOC. The reason is attributed to the non-unitarity of the Hermitian operators we use.

\end{document}